\documentclass[a4paper,onecolumn,11pt,accepted=2022-04-04]{quantumarticle}
\pdfoutput=1
\usepackage[utf8]{inputenc}
\usepackage[english]{babel}
\usepackage[T1]{fontenc}
\usepackage{amsmath}
\usepackage{hyperref}

\usepackage{tikz}
\usepackage{lipsum}

\begin{document}

\title{Environment-assisted strong coupling regime}

\author{Timofey T. Sergeev}
\affiliation{Dukhov Research Institute of Automatics (VNIIA), 22 Sushchevskaya, Moscow 127055, Russia}
\affiliation{Moscow Institute of Physics and Technology, 9 Institutskiy pereulok, Dolgoprudny 141700, Moscow region, Russia}
\affiliation{Institute for Theoretical and Applied Electromagnetics, 13 Izhorskaya, Moscow 125412, Russia}

\author{Ivan V. Vovcenko}
\affiliation{Moscow Institute of Physics and Technology, 9 Institutskiy pereulok, Dolgoprudny 141700, Moscow region, Russia}

\author{Alexander A. Zyablovsky}
\email{zyablovskiy@mail.ru}
\affiliation{Dukhov Research Institute of Automatics (VNIIA), 22 Sushchevskaya, Moscow 127055, Russia}
\affiliation{Moscow Institute of Physics and Technology, 9 Institutskiy pereulok, Dolgoprudny 141700, Moscow region, Russia}
\affiliation{Institute for Theoretical and Applied Electromagnetics, 13 Izhorskaya, Moscow 125412, Russia}
\affiliation{Kotelnikov Institute of Radioengineering and Electronics RAS, 11-7 Mokhovaya, Moscow 125009, Russia}

\author{Evgeny S. Andrianov}
\affiliation{Dukhov Research Institute of Automatics (VNIIA), 22 Sushchevskaya, Moscow 127055, Russia}
\affiliation{Moscow Institute of Physics and Technology, 9 Institutskiy pereulok, Dolgoprudny 141700, Moscow region, Russia}
\affiliation{Institute for Theoretical and Applied Electromagnetics, 13 Izhorskaya, Moscow 125412, Russia}

\maketitle

\begin{abstract}
  Strong coupling regime takes place in open hybrid systems consisting of two or more physical subsystems when the coupling strength between subsystems exceeds the relaxation rate. The relaxation arises due to the interaction of the system with environment. For this reason, it is usually believed that the enhancement of the interaction with environment inevitably leads to a transition of the system from the strong to weak coupling regime. In this paper, we refute this common opinion. We demonstrate the interaction of the coupled system with environment induces an additional coupling between the subsystems that contribute to retention the system in the strong coupling regime. We show that the environmental-induced coupling strength is proportional to the product of the Rabi coupling strength by the gradient of the density of states of the reservoir. There is a critical Rabi coupling strength above which the environmental-induced coupling ensures that the system remains in the strong coupling regime at any relaxation rate. In this case, the strong coupling regime takes place even when the relaxation rate significantly exceeds the Rabi coupling strength between the subsystems. The critical coupling depends on the gradient of the reservoir density of states. We demonstrate that managing this gradient can serve as an additional tool to control the properties of the coupled systems.
\end{abstract}

Recently, non-Hermitian systems consisting of several coupled subsystems have received significant interest \cite{ref1}-\cite{ref7}. The increase of coupling strength between the subsystems leads to a transition from the weak coupling (WC) to the strong coupling (SC) regime \cite{ref1}-\cite{ref11}. The transition between the WC and SC regimes occurs at an exceptional point (EP) \cite{ref12}-\cite{ref82}. The EP is a spectral singularity in the parameter space of a non-Hermitian system, where two or more eigenfrequencies coincide and the corresponding eigenstates become collinear \cite{ref12,ref13,ref15,ref16}. In the SC regime, hybrid eigenstates of the interacting subsystems are formed and Rabi splitting in the spectrum appears \cite{ref1}-\cite{ref5}. The formation of hybrid states leads to a change in the physical properties of such systems \cite{ref5},\cite{ref17}-\cite{ref24}, which makes them promising for different applications \cite{ref25}-\cite{ref32}. For example, the plasmonic \cite{ref1}-\cite{ref5}, photonic \cite{ref6,ref8,ref9} and polaritonic \cite{ref33-} realizations of strongly-coupled non-Hermitian systems are used to enhance the sensitivity of laser gyroscopes \cite{ref34} and sensors \cite{ref35,ref36}, to control the rate of chemical reactions \cite{ref5,ref22,ref23,ref37,ref38}, to build quantum information systems \cite{ref39}, to achieve single-mode lasing in multimode systems \cite{ref40,ref41} and lasing without inversion \cite{ref25,ref26}.

To achieve SC regime, the different ways to decrease the relaxation rate and increase the coupling strength between subsystems are used. The relaxation arises due to the interaction of the system with the environment. For this reason, it is assumed that an increase in the magnitude of the interaction with environment lead to a transition from the strong to weak coupling regime and, as a result, destroyed the hybrid states and hinders the practical applications of the coupled systems. 

We refute this opinion and demonstrate that interaction with the environment can even lead to an increase in the coupling strength between the subsystems.
From a theoretical point of view, when the subsystems strongly interact, not only is the Hermitian part of the Hamiltonian modified but also the description of the relaxation processes should be revisited \cite{ref42}-\cite{ref51}. Therefore, relaxation superoperators should be derived by considering the eigenstates of the interacting subsystems \cite{ref52}-\cite{ref54}. This results in the appearance of cross-relaxation processes when the relaxation of one subsystem depends on the state of another \cite{ref55}-\cite{ref58}. Mathematically, this manifests as the appearance of a non-Hermitian addition to the coupling strength arising from the Hamiltonian interaction. Usually, this effect is ignored when considering the SC systems.

In this paper, we demonstrate that in open coupled systems, the interaction with environment lead to appearance of a additional coupling between the subsystems (hereinafter, environmental-induced (EI) coupling). By taking account reservoir degrees of freedom, we show that environmental-induced coupling is proportional to a gradient of reservoir density of states and the Rabi coupling between subsystems. The EI coupling results in a qualitative change in the system behavior, namely, it leads to the repulsion of eigenfrequencies, which increases with both the Rabi coupling strength and the relaxation rate. In addition, the EI coupling leads to increase of the interaction energy between the subsystems. That promotes the transition to the SC regime. We show that there is a critical coupling strength, above which the EI coupling guarantees that the system is in the SC regime at any value of the relaxation rates. Thus, when the coupling strength exceeds the critical value, the strong coupling regime is resistant to destruction caused by losses. This regime can be referred to as the environment-assisted strong coupling regime (EASC regime). In the EASC regime, the hybrid states exist even when the relaxation rate is significantly above the Rabi coupling strength between the subsystems.
The critical coupling and the eigenstates of the coupled system depend on the gradient of the reservoir density of states. This opens additional possibility to control of the interaction energy between subsystems by manipulation of the reservoir density of states.

\section{Model}
We consider a system of two coupled oscillators, interacting with their reservoirs that are non-correlated. The Hamiltonian of this system is 

\begin{equation}
\label{eq1n}
  \hat H = {\hat H_S} + {\hat H_R} + {\hat H_{SR}}
\end{equation}
 where ${\hat H_S} = \omega_0 \,\hat a_1^\dag {\hat a_1} + \omega_0 \,\hat a_2^\dag {\hat a_2} + \Omega \left( {\hat a_1^\dag {{\hat a}_2} + \hat a_2^\dag {{\hat a}_1}} \right)$ is the Hamiltonian for two coupled oscillators in the rotating-wave approximation \cite{ref71,ref72}; $\hbar  = 1$. Here $\hat a_{1,2}$ and $\hat a_{1,2}^\dag$ are the annihilation and creation operators of the first and the second oscillators, respectively. $\omega_0$ is the oscillator frequency. The last term $\Omega \left( {\hat a_1^\dag {{\hat a}_2} + \hat a_2^\dag {{\hat a}_1}} \right)$ is the interaction energy between oscillators, where $\Omega $ is the Rabi coupling strength between the oscillators \cite{ref71,ref72}. This Hamiltonian is used, for example, to describe the interaction of electromagnetic modes in two coupled resonators. In this case, the last term has meaning of the interaction energy of cavity electric fields in the second quantization formalism \cite{ref71,ref72}.
 ${\hat H_R}$ and ${\hat H_{SR}}$ are the Hamiltonians of the reservoirs and the interaction of the oscillators with the reservoirs, respectively (see, for details, Appendix A).

When the oscillators are uncoupled, the interaction of the oscillators with the reservoirs leads to an independent relaxation process in each oscillator. To describe the dynamics of the coupled oscillators, one usually assumes that the oscillator relaxation rates do not depend on the coupling between them (local approach) \cite{ref59,ref60}. In this approach, the degrees of freedom of reservoirs are eliminated in the Born-Markov approximation, which leads to a master equation for the system density matrix  $\hat \rho $ in the Lindblad form \cite{ref71,ref72}. Using the equalities $\left\langle {{{\hat a}_{1,2}}} \right\rangle  = Tr\left( {{{\hat a}_{1,2}}\hat \rho } \right)$ and $d\left\langle {{{\hat a}_{1,2}}} \right\rangle /dt = Tr\left( {{{\hat a}_{1,2}}\dot{ \hat \rho }} \right)$ the closed equations for the average values of the annihilation operators $\left\langle {{{\hat a}_{1,2}}} \right\rangle $ can be obtained

\begin{align}
\label{eq1}
\frac{d}{{dt}}\left( {\begin{array}{*{20}{c}}
{{a_1}}\\
{{a_2}}
\end{array}} \right) = \left( {\begin{array}{*{20}{c}}
{ - i\,{\omega _0} - {\gamma _1}}&{ - i\Omega }\\
{ - i\Omega }&{ - i\,{\omega _0} - {\gamma _2}}
\end{array}} \right)\left( {\begin{array}{*{20}{c}}
{{a_1}}\\
{{a_2}}
\end{array}} \right)
\end{align}
where ${a_{1,2}} = \left\langle {{{\hat a}_{1,2}}} \right\rangle$ are amplitudes of the first and the second oscillator, ${\gamma _{1,2}}$ are the relaxation rates of the oscillators. The eigenfrequencies of this system are ${\omega _{1,2}} = {\omega _0} - i\frac{{{\gamma _1} + {\gamma _2}}}{2} \pm \sqrt {{\Omega ^2} - {{\left( {{\gamma _1} - {\gamma _2}} \right)}^2}/4}$ and the eigenstates are 
\begin{align}
\label{eq2}
{\vec e_{1,2}} = {\left( {\begin{array}{*{20}{c}}
{{a_1}}&{{a_2}}
\end{array}} \right)^T} = {\left( {\begin{array}{*{20}{c}}
{i\left( {{\gamma _2} - {\gamma _1} + \sqrt {{{\left( {{\gamma _1} - {\gamma _2}} \right)}^2} - 4{\Omega ^2}} } \right)/2\Omega ,}&1
\end{array}} \right)^T}
\end{align}
 \cite{ref12,ref13}.
 
This system has an exceptional point (EP), at which the eigenstates are linearly dependent and the eigenfrequencies coincide. The EP takes place when the coupling strength of the oscillators equals $\Omega  = {\Omega _{EP}} = \left| {{\gamma _1} - {\gamma _2}} \right|/2$ \cite{ref12,ref13}. If $\Omega  < {\Omega _{EP}}$, the WC regime takes place in the system. In this regime, the real parts of the eigenfrequencies coincide with each other and there is an exponential decay of the oscillators amplitudes with time \cite{ref12,ref13,ref82}. Also, we determine the interaction energy for each of the eigenstates, i.e., the interaction energy in the system when the system is in one of the eigenstates. This interaction energy is calculated by the formula $\Omega \left( {a_1^*{a_2} + {a_1}a_2^*} \right)$, where $a_1$ and $a_2$ are elements of the corresponding eigenstate~(\ref{eq2}). In the WC regime, the interaction energy between the oscillators is zero for both the eigenstates. If $\Omega  > {\Omega _{EP}}$, the SC regime takes place in the system. In the SC regime, the real parts of the eigenfrequencies are split \cite{ref12,ref13,ref82} and the exponential decay is replaced by decaying oscillations of the oscillators amplitudes. In the SC regime, the interaction energy between the oscillators is nonzero for both the eigenstates. It is negative for the eigenstate with lower frequency and positive for the eigenstate with higher frequency. Thus, the EP separates the WC and SC regimes in the system of the coupled oscillators. Note that the exceptional points exist not only in the equations for average amplitudes, but also in the master equations for the density matrix  of the systems with asymmetrical relaxation rates \cite{ref78}-\cite{ref81}. In particular, they are studied in the language of the generalized master equations \cite{ref77,ref78,ref84} that take into account the incoherent coupling between modes.

The assumption regarding the independence of the relaxation rates of the oscillators is not valid when the coupling strength between subsystems are comparable or greater than the relaxation rates \cite{ref52}-\cite{ref54}. This is because the eigenstates of the system of coupled oscillators do not coincide with the eigenstates of the non-interacting oscillators. It is necessary to eliminate the reservoir degrees of freedom using the eigenstates of the interacting subsystems \cite{ref52}-\cite{ref54}.

To derive the consistent equations describing the system dynamics in both the WC and SC regimes, we eliminate the reservoirs degrees of freedom using the partial-secular approximation (see Appendices A and B) \cite{ref58}. The obtained equations for density matrix include additional terms, which describe the cross-relaxation of the eigenstates (see Appendices A and B and also \cite{ref77}). Then using the equalities $\left\langle {{{\hat a}_{1,2}}} \right\rangle  = Tr\left( {{{\hat a}_{1,2}}\hat \rho } \right)$ and $d\left\langle {{{\hat a}_{1,2}}} \right\rangle /dt = Tr\left( {{{\hat a}_{1,2}}\dot{ \hat \rho }} \right)$ we derive the following closed equations for the average values of the annihilation operators $\left\langle {{{\hat a}_{1,2}}} \right\rangle $

\begin{align}
\label{eq3}
\begin{array}{l}
\frac{d}{{dt}}\left( {\begin{array}{*{20}{c}}
{{a_1}}\\
{{a_2}}
\end{array}} \right) = \left( {\begin{array}{*{20}{c}}
{ - i\,{\omega _0} - {\gamma _1}}&{ - i\Omega  - {\rm K}\left( {\Omega ,\,{\gamma _1}} \right)}\\
{ - i\Omega  - {\rm K}\left( {\Omega ,\,{\gamma _2}} \right)}&{ - i\,{\omega _0} - {\gamma _2}}
\end{array}} \right)\left( {\begin{array}{*{20}{c}}
{{a_1}}\\
{{a_2}}
\end{array}} \right)
\end{array}
\end{align}
where the relaxation rates ${\gamma _{1,2}}\left( \omega  \right) = \pi {\left| {\gamma _\omega ^{(1),(2)}} \right|^2}\rho \left( \omega  \right)$ are determined by the reservoir densities of states $\rho \left( \omega  \right)$ and the interaction constant $\gamma _\omega ^{(1),(2)}$ of the oscillators with its reservoirs. The functions ${\rm K}\left( {\Omega ,\,{\gamma _{1,2}}} \right)$ are an additional coupling strengths, which appear due to the interaction of the coupled system with the environment (environmental-induced (EI) coupling). These terms appear in both classical and quantum considerations and are determined by the frequency dispersion of the reservoir density of states (see Appendices A and B),

\begin{align}
\label{eq4}
\begin{array}{l}
{\rm K}\left( {\Omega ,\,{\gamma _{1,2}}} \right) = \frac{{{\gamma _{1,2}}({\omega _s}) - {\gamma _{1,2}}({\omega _a})}}{2} \approx \frac{{\pi {{\left| {\gamma _{{\omega _0}}^{(1),(2)}} \right|}^2}}}{2}{\left. {\frac{{\partial \rho \left( \omega  \right)}}{{\partial \omega }}} \right|_{{\omega _0}}}\left( {{\omega _s} - {\omega _a}} \right)
\end{array}
\end{align}
where ${\omega _s}$ and ${\omega _a}$ are the eigenfrequencies of the symmetric and anti-symmetric states of the Hermitian system, respectively. Note that in the case of frequency-independent the reservoir density of states the EI coupling strength is zero.

We emphasize that due to we consider the system of two bosonic modes (oscillators) with quadratic interaction~\eqref{eq1n} Eq.~\eqref{eq3} for average amplitudes is derived from the master equations without additional assumptions \cite{ref78}. Therefore, the dynamics of $\left\langle {\hat a_1} \right\rangle $ and $\left\langle {\hat a_2} \right\rangle $ calculated from Eq.~\eqref{eq3} coincides with one obtained from the master equations for density matrix \cite{ref78} (see Appendices A and C). 

\section{Formation of environment-assisted strong coupling regime}
The EI coupling leads to an increase in the interaction between the oscillators. Therefore, it can be expected that due to the presence of EI coupling, the transition between the WC and SC regimes occurs at smaller values of the Rabi coupling strength or at greater vales of the relaxation rates. To be specific, we consider the case of a power-dependent dispersion of the density of states ($\rho \left( \omega  \right) \sim {\omega ^n}$). Then

\begin{align}
\label{eq5}
{\rm K}\left( {\Omega ,\,{\gamma _{1,2}}} \right) \approx n\,\Omega \,{\gamma _{1,2}}({\omega _0})/{\omega _0}
\end{align}
and the eigenfrequencies of the oscillators system Eq.~\eqref{eq3} take the form

\begin{align}
\label{eq6}
\begin{array}{l}
{\omega _{1,2}} = {\omega _0} - i\frac{{{\gamma _1} + {\gamma _2}}}{2} \pm \sqrt {{\Omega ^2}\left( {1 - \frac{{{n^2}{\gamma _1}{\gamma _2}}}{{\omega _0^2}} - i\,\frac{{n\left( {{\gamma _1} + {\gamma _2}} \right)}}{{{\omega _0}}}} \right) - {{\left( {{\gamma _1} - {\gamma _2}} \right)}^2}/4} 
\end{array}
\end{align}
and experience a splitting. Because the EI coupling strength is proportional to the product of the coupling strength $\Omega$ and the relaxation rates ${\gamma _{1,2}}$ (see Eq.~\eqref{eq5}), the increase in both the coupling strength and the relaxation rates leads to a growth in the frequency splitting. This takes place both below and above the EP. In addition, the EI coupling leads to a break of the exact symmetry among eigenmodes in the SC regime (${\left| {{a_1}} \right|^2} \ne {\left| {{a_2}} \right|^2}$) and to an appearance of a nonzero interaction energy between the oscillators in the WC regime.

\begin{figure}[t]
  \centering
 \includegraphics[width=0.7\linewidth]{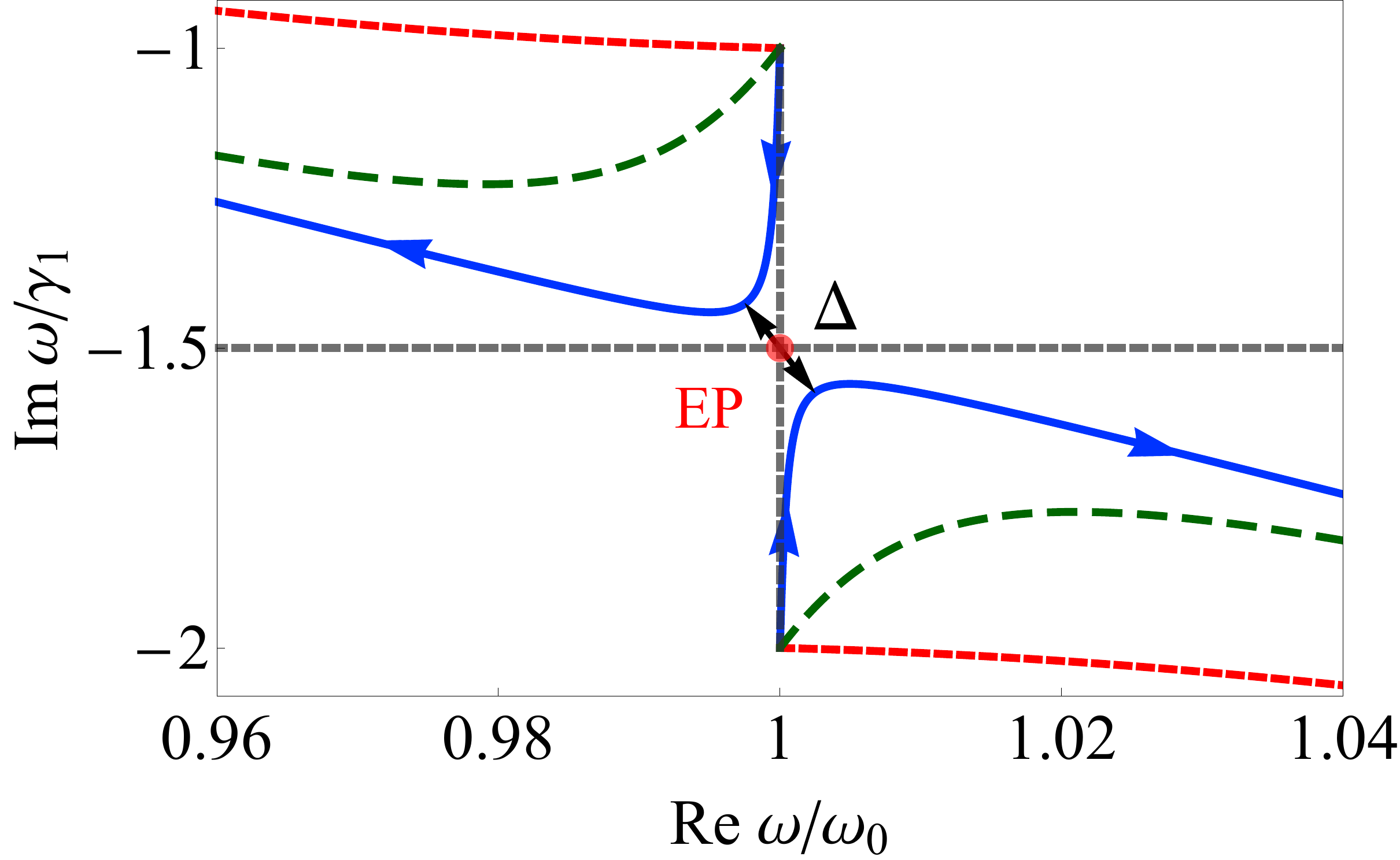}
  \caption{Trajectories of the eigenfrequencies in the complex frequency plane when the coupling strength $\Omega$ changes from $0$ to $0.1\,{\omega _0}$. The different curves are corresponded to the different values of relaxation rates: ${\gamma _1} = 2 \times {10^{ - 2}}\,{\omega _0}$ and ${\gamma _2} = {10^{ - 2}}\,{\omega _0}$ (solid blue lines), ${\gamma _1} = {10^{ - 1}}\,{\omega _0}$ and ${\gamma _2} = 5 \times {10^{ - 2}}\,{\omega _0}$ (dashed green lines), ${\gamma _1} = 6 \times {10^{ - 1}}\,{\omega _0}$ and ${\gamma _2} = 3 \times {10^{ - 1}}\,{\omega _0}$ (dotted red lines). The dashed gray line in Fig.~\ref{fig:figure1} shows the dependence when the EI coupling is not taken into account (or in the case of frequency-independent dispersion of the reservoir density of states). The red point indicates the EP in the case when the EI coupling is not taken into account.}
  \label{fig:figure1}
\end{figure}

To illustrate the influence of the EI coupling on the system behavior, we track the dependence of the eigenfrequencies ${\omega _1}$, ${\omega _2}$ (see Eq.~\eqref{eq6}) in the complex frequency plane on the coupling strength $\Omega$ (Fig.~\ref{fig:figure1}). When the EI coupling is not taken into account, the eigenfrequencies coalesce in the complex plane at a coupling strength corresponding to the EP, $\Omega  = {\Omega _{EP}}$ (Fig.~\ref{fig:figure1}). The exceptional point separates the WC ($\Omega  < {\Omega _{EP}}$) and the SC ($\Omega  > {\Omega _{EP}}$) regimes.

With taking into account the EI coupling, the eigenfrequencies do not coalesce for any values of the coupling strength (Fig.~\ref{fig:figure1}). However, there is a value of coupling strength, at which the distance $\Delta$ between the eigenfrequencies in the complex plane ($\Delta  = \left| {{\omega _1} - {\omega _2}} \right|$) reaches minimum (see designations in Fig.~\ref{fig:figure1}). Using the analogy with the case without the EI coupling, we determine this value of the coupling strength as the transition point between the WC and the SC regimes.

\begin{figure}[t]
  \centering
 \includegraphics[width=0.7\linewidth]{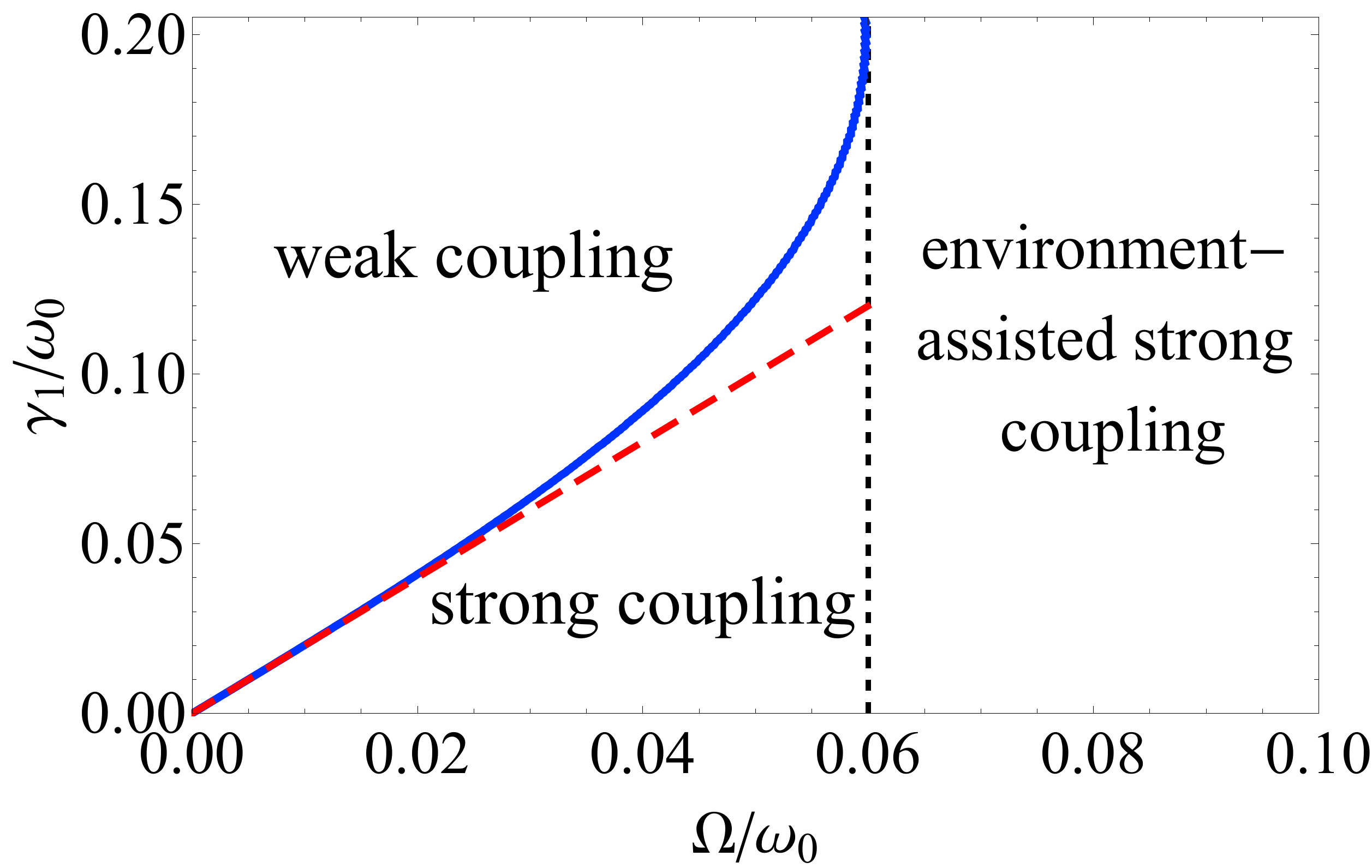}
  \caption{Phase diagram of coupling regimes in the coordinates $\Omega /{\omega _0}$ and ${\gamma _1}/{\omega _0}$ (the ratio ${\gamma _2}/{\gamma _1} = 2$ is fixed). The solid blue line shows the condition when the distance between the eigenfrequencies in the complex plane, i.e., $\Delta  = \left| {{\omega _1} - {\omega _2}} \right|$, have a minimum. The dashed red line shows the condition $\Omega  = {\Omega _{EP}} = \left| {{\gamma _1} - {\gamma _2}} \right|/2$. The vertical black line shows the critical coupling strength. The reservoir density of states $\rho \left( \omega  \right) \sim {\omega ^2}$.}
  \label{fig:figure2}
\end{figure}

In Fig.~\ref{fig:figure2}, we plot the phase diagram of different coupling regimes in the coordinates $\Omega /{\omega _0}$ and ${\gamma _1}/{\omega _0}$ (we fix the ratio ${\gamma _2}/{\gamma _1}$). When $\Omega ,{\gamma _{1,2}} \ll {\omega _0}$, the transition between the WC and SC regimes occurs at the line determined by the condition $\Omega  = {\Omega _{EP}} = \left| {{\gamma _1} - {\gamma _2}} \right|/2$ (Fig.~\ref{fig:figure2}). With the increase in the coupling strength, the transition point deviates from the condition $\Omega  = {\Omega _{EP}}$ (Fig.~\ref{fig:figure2}) and the transition between the WC and SC regimes occurs under a different condition from the one of the EP. This is due to the EI coupling strength being proportional to the product of the relaxation rate and the Rabi coupling strength. There are a critical coupling strength ${\Omega _{CP}}$ starting of which the EI coupling guarantees that the system is in the SC regime at any relaxation rate (Fig.~\ref{fig:figure2}). In this case, the increase in the relaxation rate leads to the growth of the EI coupling strength, which prevents the transition from the SC to the WC regime (see Figs.~\ref{fig:figure1} and ~\ref{fig:figure2}). Thus, when $\Omega  > {\Omega _{CP}}$, the system is in the SC regime at all values of the relaxation rate. This regime can be referred to as the environment-assisted strong coupling (EASC) regime (see Fig.~\ref{fig:figure2}).

\section{Influence of the reservoir density of states on the system states}
The EI coupling appears due to the difference in the eigenstates and eigenfrequencies of the system of coupled and uncoupled oscillators. Due to the density of states of the reservoirs depending on the frequency, the interaction with reservoirs depends on the eigenfrequencies of the coupled system \cite{ref52}-\cite{ref54}. These eigenfrequencies change when the coupling strength between the oscillators increases. As a result, the relaxation terms turn out to be explicitly dependent on the coupling strength between the oscillators and the rate of amplitude relaxation in the first/second oscillator depends on the amplitude of the second/first oscillator, which corresponds to the appearance of the EI coupling strength.

\begin{figure}[t]
  \centering
  \includegraphics[width=1\linewidth]{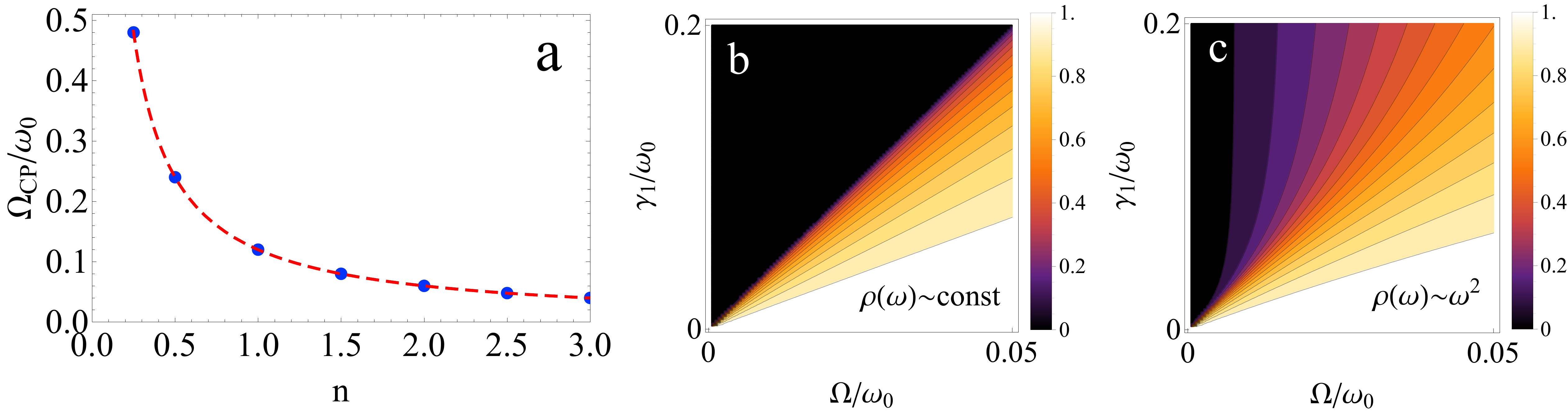}
  \caption{(a) Dependence of critical coupling strength ${\Omega _{CP}}$ on the degree $n$ in the power-law dependence of the reservoir density of states on frequency ($\rho \left( \omega  \right) \sim {\omega ^n}$). (b), (c) Diagrams of the interaction energy between the oscillators ($\Omega \left( {a_1^*{a_2} + {a_1}a_2^*} \right)$) calculated for the eigenstate amplitudes. The diagrams for other eigenstate are similar. The ratio ${\gamma _2}/{\gamma _1} = 2$ is fixed. (b) $\rho \left( \omega  \right)$ is frequency-independent, (c) $\rho \left( \omega  \right) \sim {\omega ^2}$.}
  \label{fig:figure3}
\end{figure}

The EI coupling strength depends on the frequency dispersion of the reservoir density of states (see Eq.~\eqref{eq4}). The dependence of ${\rm K}\left( {\Omega ,\,{\gamma _{1,2}}} \right)$ on the reservoir density of states leads to the dependence of the critical coupling strength ${\Omega _{CP}}$, the eigenfrequencies and the eigenstates on the same quantity (see Fig.~\ref{fig:figure3}). Let us consider the case of power-dependent density of states, $\rho \left( \omega  \right) \sim {\omega ^n}$. In the case of reservoirs with a frequency-independent density of states, $n = 0$, the EI coupling strength ${\rm K}\left( {\Omega ,\,{\gamma _{1,2}}} \right)$ is zero and does not causes the frequency splitting (see dashed gray lines in Fig.~\ref{fig:figure1}). However, in the case of reservoirs with a frequency-dependent density of states, the EI coupling leads to the frequency splitting (see Fig.~\ref{fig:figure1}). The stronger the dependence of density of states on the frequency, i.e., the higher $n$, the larger the EI coupling strength.

In addition, the EI coupling leads to a change in the eigenstates. In particular, it results in an increase of the interaction energy between the oscillators, i.e., $\Omega \left( {a_1^*{a_2} + {a_1}a_2^*} \right)$, calculated for the eigenstate amplitudes. Without taking into account the EI coupling or in the case of reservoirs with a frequency-independent density of states, in the WC regime, the interaction energy between the oscillators is zero (Fig.~\ref{fig:figure3}b). In contrast, in the case of reservoirs with a frequency-dependent density of states, the interaction energy is nonzero at any values of the coupling strength and the relaxation rates (Fig.~\ref{fig:figure3}c). That is, due to the EI coupling, the eigenstates become more coupled.

Thus, we conclude that the control of the reservoir density of states can serve as an additional tool to manipulate the eigenfrequencies and the eigenstates of the coupled systems. This control can be achieved by using the different types of reservoirs. For example, the radiation in the one-dimensional waveguide corresponds to the interaction with the reservoir having a frequency-independent density of states. While, the interaction with phonons in the solid state corresponds to the interaction with the reservoir having the square power-law dependence of the reservoir density of states on frequency. The artificial reservoirs with high gradient of the density of states can be used to decrease the critical coupling strength ${\Omega _{CP}}$. The stopped-light waveguides \cite{ref76} serve as an example of such reservoirs. In these waveguides the density of states has sharp peak near the cutoff frequency \cite{ref76} that corresponds to the gradient of the density of states ${\left. {\partial \rho /\partial \omega \,} \right|_{\omega  = {\omega _0}}} \sim {10^3}\rho \left( {{\omega _0}} \right)/{\omega _0}$. Usage the reservoirs with such gradients of the density of state can allow to decrease critical coupling strength ${\Omega _{CP}}$ to ${10^{ - 4}}{\omega _0}$.

\section{Influence of the EI coupling on the system dynamics}

\begin{figure}[t]
  \centering
  \includegraphics[width=0.6\linewidth]{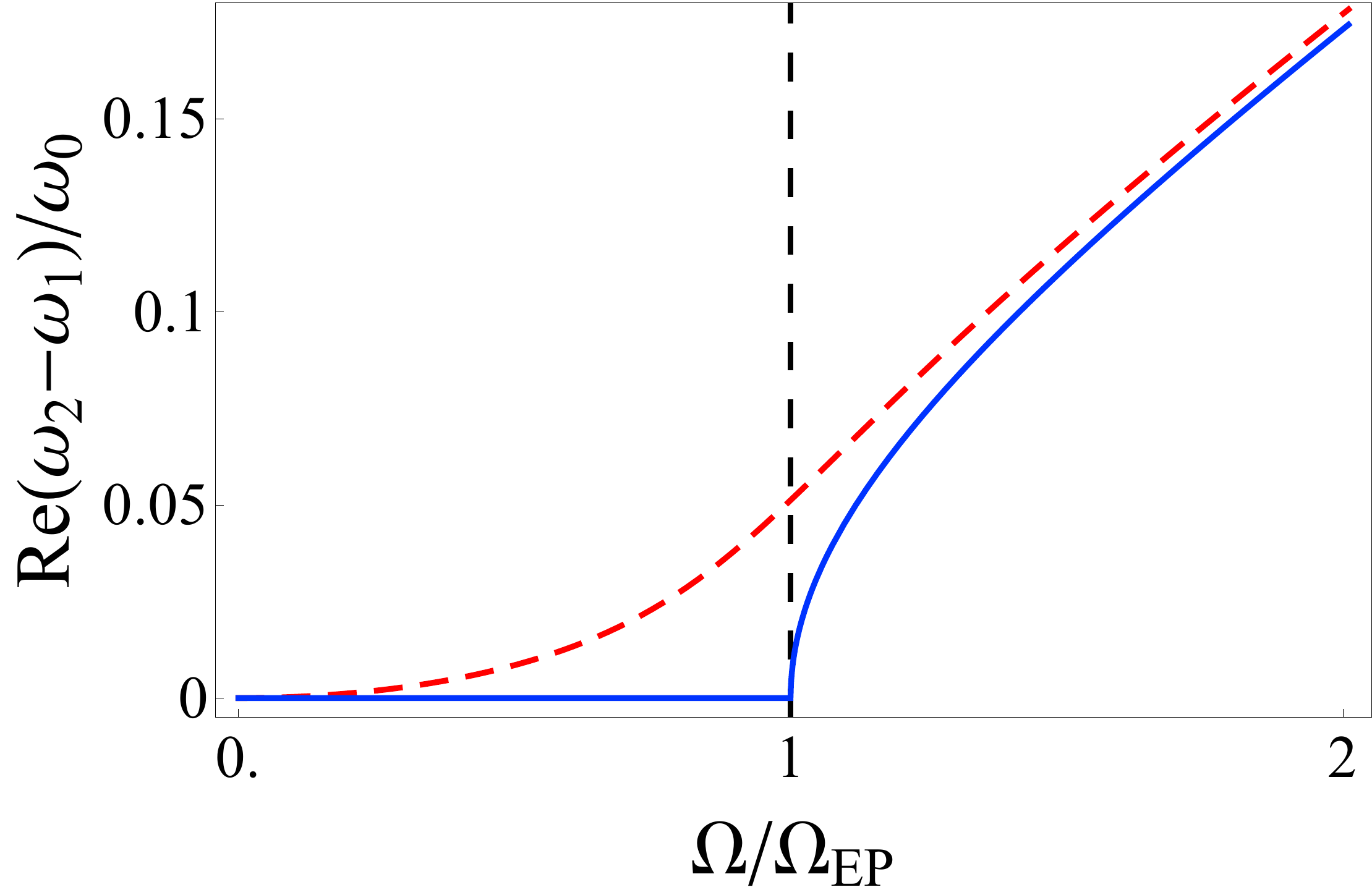}
  \caption{Dependence of the difference of the real parts of the eigenfrequencies on the coupling strength when $\rho \left( \omega  \right)$ is frequency-independent (solid blue line) and $\rho \left( \omega  \right) \sim {\omega ^2}$ (dashed red line). The vertical dashed line shows the transition between the WC and the SC regimes in the system with the frequency-independent reservoir density of states.  $\gamma_1=0.01 \omega_0$; $\gamma_1=0.02 \omega_0$; ${\Omega _{EP}} = \left| {{\gamma _1} - {\gamma _2}} \right|/2$.}
  \label{fig:figure4}
\end{figure}

The EI coupling leads to changing in the eigenstates and eigenfrequencies, which, in turn, causes changes in the system dynamics. Firstly, the EI coupling leads to increase in the difference of the real part of eigenfrequencies (Fig.~\ref{fig:figure4}) that causes growth of the oscillations frequency of the average amplitudes $\left\langle {{{\hat a}_1}} \right\rangle$ and $\left\langle {{{\hat a}_2}} \right\rangle $. Such a change is noticeable in the WC regime (Fig.~\ref{fig:figure4}), where the EI coupling promotes the transition to the SC regime.

Secondly, the EI coupling leads to a qualitative change in the temporal dependence of the interaction energy. As an example, we simulate the time dynamics of two coupled oscillators, when at the initial time the first oscillator is in the excited state and the second oscillator is in the ground state. Our calculations based on Eq.~\eqref{eq3} show that in the case of frequency-independent density of states (the EI coupling strength is zero) the interaction energy remains zero at all times. It is due to that in the SC regime when $K\left( {\Omega ,\,{\gamma _{1,2}}} \right) = 0$ ($\rho \left( \omega  \right) = const$) the relaxation rates of two eigenmodes are equal to each other. So long as the initial amplitudes of two eigenstates are the same and the interaction energies of eigenstates differ in the sign, the total interaction energy stays zero throughout the system evolution. In the WC regime, the interaction energies for each of the eigenstates are zero (Fig.~\ref{fig:figure3}b, c) and so the total interaction energy stays zero throughout the system evolution too.

\begin{figure}[t]
  \centering
  \includegraphics[width=0.6\linewidth]{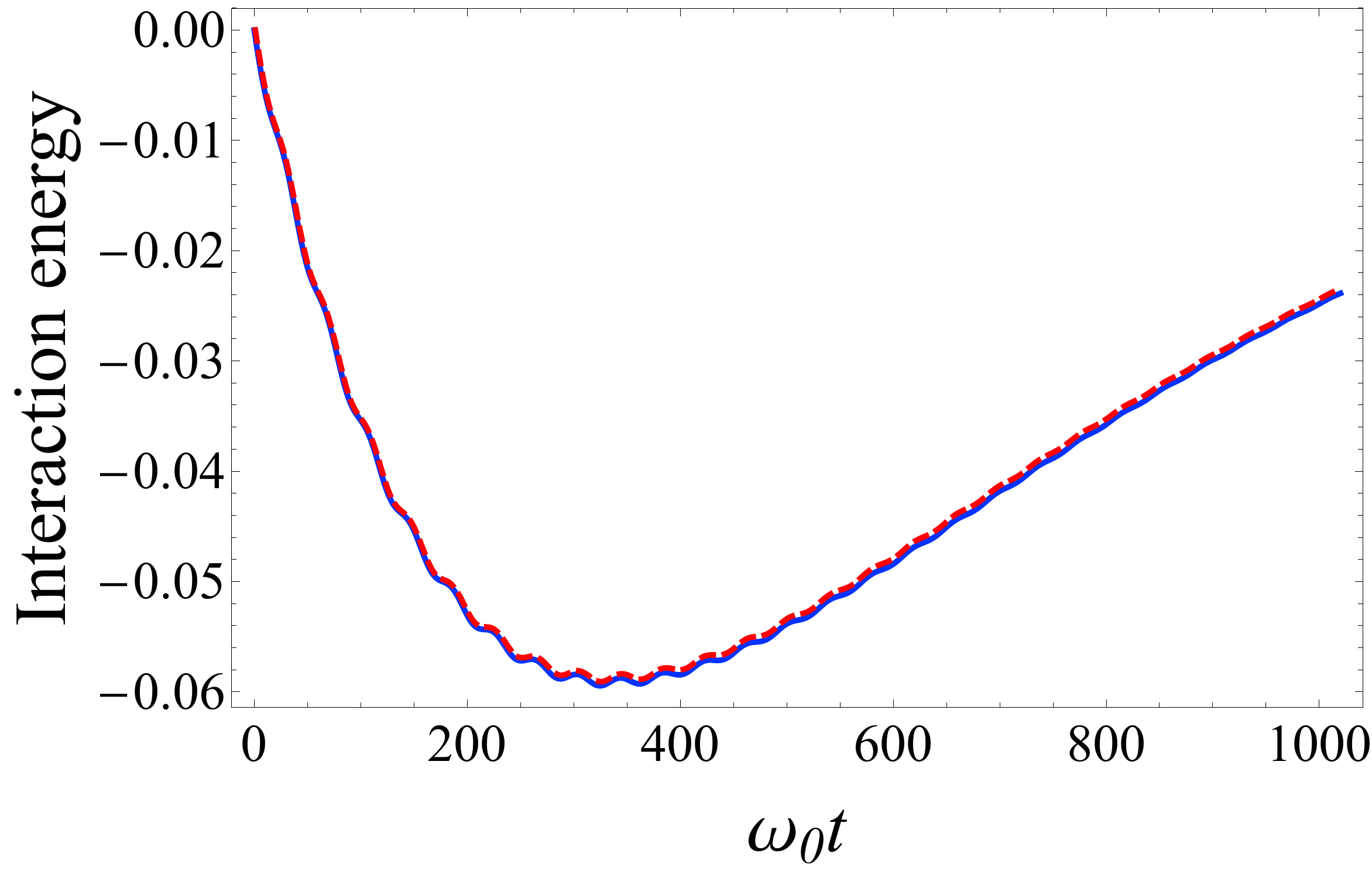}
  \caption{Dependence of the interaction energy calculated by Eq.~\eqref{eq3} $\left( {a_1^*{a_2} + {a_1}a_2^*} \right)$ (solid blue line) and by the master equation for density matrix $\left\langle {\hat a_1^\dag {{\hat a}_2} + {{\hat a}_1}\hat a_2^\dag } \right\rangle$ (dashed red line) in the EASC regime. $\rho \left( \omega  \right) \sim {\omega ^2}$; $\gamma_1 = 0.001 \omega_0$; $\gamma_2 = 0.002 \omega_0$; the Rabi coupling strength $\Omega = 0.08 \omega_0$.}
  \label{fig:figure5}
\end{figure}

In the case of reservoirs with a frequency-dependent density of states, the interaction energy differs from zero (Fig.~\ref{fig:figure5}). It is due to the two factors. First, the EI coupling leads to the difference in the relaxation rates of two eigenmodes for all Rabi coupling strength (Fig.~\ref{fig:figure1}). As a result, at $t >  > {\left| {{\gamma _1} - {\gamma _2}} \right|^{ - 1}}$ the system evolves to a state in which one of the eigenstates dominates. In turn, the interaction energy for each of the eigenstates is not zero (Figs.~\ref{fig:figure3}c). For this reason, the interaction energy differs from zero during the system evolution from the excited state (Fig.~\ref{fig:figure5}). That is, the EI coupling leads to a non-zero interaction energy during the system evolution.

Note that the interaction energy $\Omega \left\langle {\hat a_1^\dag {{\hat a}_2} + {{\hat a}_1}\hat a_2^\dag } \right\rangle $ as well as the energies of the oscillators ${\omega _0}\left\langle {\hat a_1^\dag {{\hat a}_1}} \right\rangle $ and ${\omega _0}\left\langle {\hat a_2^\dag {{\hat a}_2}} \right\rangle $ can be calculated from Eq.~\eqref{eq3} only approximately by using the semi-classical replacement, such as $\left\langle {\hat a_1^ + {{\hat a}_2} + {{\hat a}_1}\hat a_2^ + } \right\rangle  \to \left\langle {\hat a_1^ + } \right\rangle \left\langle {{{\hat a}_2}} \right\rangle  + \left\langle {{{\hat a}_1}} \right\rangle \left\langle {\hat a_2^ + } \right\rangle $. To verify our predictions for the quantum system, we simulate dynamics of two coupled oscillators by the master equations for density matrix (see Appendix C). We compare the system dynamics calculated from Eq.~\eqref{eq3} and from the master equations for the density matrix~\eqref{eq41} for the different reservoir density of states (Appendix C). It is seen that the master equation for the density matrix (dashed red line in Fig.~\ref{fig:figure5}) predicts a similar system dynamics as Eq.~\eqref{eq3} (solid blue line in Fig.~\ref{fig:figure5}).

Thus, we conclude that the EI coupling significantly influence on the system dynamics in both semiclassical and quantum cases. This opens the way to control the quantum states of the system by the reservoir density of states.

\section{Relation between environment-assisted strong coupling and ultra-strong coupling regimes}
The environment-assisted strong coupling regime takes place when the coupling strength exceeds the critical value. This value depends on the reservoir density of states and can be comparable to the eigenfrequencies of the system (see Fig.~\ref{fig:figure3}). It is known \cite{ref61,ref62} that an increase in the coupling strength leads to a transition from the SC to the ultra-strong coupling (USC) regime in which the counter-rotating and the diamagnetic terms in the Hamiltonian begin to have a noticeable effect on the system states \cite{ref63}-\cite{ref65}. 

Usually, it is considered that the transition to the USC regime occurs when the coupling strength reaches one tenth of the system eigenfrequency \cite{ref61,ref63},\cite{ref66}-\cite{ref69} (see also Appendix C). Depending on the reservoir density of states, the transition to the EASC regime can precede the transition to the USC regime and vice versa. In the first case, an increase in the coupling strength, at the beginning, leads to formation of the hybrid states at all values of the relaxation rates and the system goes into the EASC regime. At further increase in the coupling strength, the counter rotating and the diamagnetic terms begin to modify these hybrid states and the transition to the USC regime occurs. In the second case, the transition to the USC regime occurs without the intermediate transition to the EASC regime. Thus, we conclude that depending on the reservoir density of states, there are two possible types of transition to the USC regime.

\section{Summary}
To conclude, we consider a strong coupled system of two oscillators interacting with their reservoirs. We demonstrate that the interaction of the coupled oscillators with the environment leads to the appearance of additional coupling between the oscillators (environmental-induced (EI) coupling). The EI coupling results in a qualitative change in the behavior of open coupled systems. Due to the presence of the EI coupling, the transition between the WC and SC regimes occurs at smaller values of the Rabi coupling strength or at greater vales of the relaxation rates. Since the EI coupling strength is proportional to the relaxation rates, the increase in relaxation rate leads to growth of the EI coupling, which prevents the transition from the SC to the WC regime. We demonstrate that there is a critical Rabi coupling strength ${\Omega _{CP}}$ above of which the EI coupling makes it impossible to transition from the SC to the WC regime with the increase in the relaxation rates. Thus, when $\Omega  > {\Omega _{CP}}$ the system is in the SC regime at all values of the relaxation rate. This regime we define as the environment-assisted strong coupling (EASC) regime. The EI coupling depends on the gradient of density of states in the reservoirs, which leads to the corresponding dependence of the critical coupling strength, eigenfrequencies and eigenstates of the coupled system. This opens the additional way to control the behavior of the coupled systems with the reservoir density of states.

\section*{Acknowledgments}
The study was financially supported by a grant from Russian Science Foundation (project No. 20-72-10057). E.S.A. thanks foundation for the advancement of theoretical physics and mathematics “Basis”.

\appendix

\section{Master equation and equations for average amplitudes for the coupled oscillators}
Here we derive the dynamic equations for averaged amplitudes of two coupled harmonic oscillators in quantum case. Total Hamiltonian of two coupled oscillators interacting with the reservoirs has the form
\begin{equation}
\label{eq7}
  \hat H = {\hat H_S} + {\hat H_R} + {\hat H_{SR}}
\end{equation}
 where
\begin{equation}
\label{eq8}
  {\hat H_S} = \omega_0 \,\hat a_1^\dag {\hat a_1} + \omega_0 \,\hat a_2^\dag {\hat a_2} + \Omega \left( {\hat a_1^\dag {{\hat a}_2} + \hat a_2^\dag {{\hat a}_1}} \right)
\end{equation}
 
\begin{equation}
\label{eq9}
{\hat H_{SR}} = \lambda \left( {\sum\limits_k {\gamma _k^{(1)}} (\hat a_1^\dag  + {{\hat a}_1})(\hat b_{1k}^\dag  + {{\hat b}_{1k}}) + \sum\limits_k {\gamma _k^{(2)}} (\hat a_2^\dag  + {{\hat a}_2})(\hat b_{2k}^\dag  + {{\hat b}_{2k}})} \right) = \lambda \left( {{{\hat S}_1}{{\hat R}_1} + {{\hat S}_2}{{\hat R}_2}} \right)
\end{equation}

\begin{equation}
\label{eq10}
{\hat H_R} = {\hat H_{{R_1}}} + {\hat H_{{R_2}}},\,\,\,\,\,\,\,\,\,\,\,\,\,\,\,\,{\hat H_{{R_1}}} = \sum\limits_k {\omega _k^{(1)}b_{1k}^\dag {b_{1k}},\,\,\,\,\,\,\,\,\,\,\,\,\,\,\,\,} {\hat H_{{R_2}}} = \sum\limits_k {\omega _k^{(2)}b_{2k}^\dag {b_{2k}}}
\end{equation}
Here $\omega_0$ is the oscillator eigenfrequency, $\Omega $ is the interaction constant between them, $\omega _k^{(1)}$ and $\omega _k^{(2)}$ are reservoir mode eigenfrequencies, $\gamma _k^{\left( 1 \right)}$ and $\gamma _k^{\left( 2 \right)}$ are the interaction constants between oscillators and their reservoirs.
 
In the interaction representation the von Neumann equation for the density matrix has the form

\begin{equation}
\label{eq11}
\frac{{\partial \hat {\tilde \rho} }}{{\partial t}} = i\,[\hat {\tilde \rho} ,{\hat {\tilde H}_{SR}}]
\end{equation}
where
\begin{equation}
\label{eq12}
{\hat {\tilde H}_{SR}} = \lambda \,({\hat {\tilde S}_1}{\hat {\tilde R}_1} + {\hat {\tilde S}_2}{\hat {\tilde R}_2})
\end{equation}

\begin{equation}
\label{eq13}
{\hat {\tilde S}_1} = \exp (i{\hat {\tilde H}_S}t){\hat S_1}\exp ( - i{\hat {\tilde H}_S}t)
\end{equation}

\begin{equation}
\label{eq14}
{\hat {\tilde R}_1} = \exp (i{\hat {\tilde H}_R}t){\hat R_1}\exp ( - i{\hat {\tilde H}_R}t)
\end{equation}

\begin{equation}
\label{eq15}
{\hat {\tilde S}_2} = \exp (i{\hat {\tilde H}_S}t){\hat S_2}\exp ( - i{\hat {\tilde H}_S}t)
\end{equation}

\begin{equation}
\label{eq16}
{\hat {\tilde R}_2} = \exp (i{\hat {\tilde H}_R}t){\hat R_2}\exp ( - i{\hat {\tilde H}_R}t)
\end{equation}
We will consider $\lambda $ as small parameter and expand density matrix in the series of $\lambda $,

\begin{equation}
\label{eq17}
\hat {\tilde \rho} (t) = {\hat {\tilde \rho} _0}(t) + \lambda {\hat {\tilde \rho} _1}(t) + {\lambda ^2}{\hat {\tilde \rho} _2}(t) + O({\lambda ^3})
\end{equation}
In zero, first and second orders we have the following equations
\begin{equation}
\label{eq18}
\frac{{\partial {{\hat {\tilde \rho} }_0}(t)}}{{\partial t}} = 0,\,\,\,\,\frac{{\partial {{\hat {\tilde \rho} }_1}(t)}}{{\partial t}} = i\,[{\hat {\tilde \rho} _0}(t),{\hat {\tilde S}_1}{\hat {\tilde R}_1} + {\hat {\tilde S}_2}{\hat {\tilde R}_2}],\,\,\,\,\frac{{\partial {{\hat {\tilde \rho} }_2}(t)}}{{\partial t}} = i\,[{\hat {\tilde \rho} _1}(t),{\hat {\tilde S}_1}{\hat {\tilde R}_1} + {\hat {\tilde S}_2}{\hat {\tilde R}_2}]
\end{equation}
with the initial conditions $\hat {\tilde \rho} ({t_0}) = {\hat {\tilde \rho} _0}({t_0})$, ${\hat {\tilde \rho}_1}({t_0}) = 0$, ${\hat {\tilde \rho} _2}({t_0}) = 0$. Formal integration of Eq.~\eqref{eq18} gives:

\begin{equation}
\label{eq19}
{\hat {\tilde \rho} _0}({t_0} + \Delta t) = \hat {\tilde \rho} \left( {{t_0}} \right)
\end{equation}

\begin{equation}
\label{eq20}
{\hat {\tilde \rho} _1}({t_0} + \Delta t) = \\ i\int\limits_{{t_0}}^{{t_0} + \Delta t} {d{t_1}} \left[ {{{\hat {\tilde \rho} }_0}({t_0}),{{\hat {\tilde S}}_1}({t_1}){{\hat {\tilde R}}_1}({t_1}) + {{\hat {\tilde S}}_2}({t_1}){{\hat {\tilde R}}_2}({t_1})} \right]
\end{equation}

\begin{equation}
\label{eq21}
\begin{array}{l}
{{\hat {\tilde \rho} }_2}({t_0} + \Delta t) = \\
 - \int\limits_{{t_0}}^{{t_0} + \Delta t} {d{t_1}} \left[ {\int\limits_{{t_0}}^{{t_1}} {d{t_2}[{{\hat {\tilde \rho} }_0}({t_0}),{{\hat {\tilde S}}_1}({t_2}){{\hat {\tilde R}}_1}({t_2}) + {{\hat {\tilde S}}_2}({t_2}){{\hat {\tilde R}}_2}({t_2})],{{\hat {\tilde S}}_1}({t_1}){{\hat {\tilde R}}_1}({t_1}) + {{\hat {\tilde S}}_2}({t_1}){{\hat {\tilde R}}_2}({t_1})} } \right]
\end{array}
\end{equation}
Further, we consider noncorrelated reservoirs. Following to the Born approximation \cite{ref71,ref72}, we present the density matrix in the form $\hat {\tilde \rho} (t) = {\hat {\tilde \rho} _S}(t)\,\hat {\tilde \rho} _{{R_1}}^{th}\,\hat {\tilde \rho} _{{R_2}}^{th}$, where superscript “th” means that corresponding reservoir is at thermodynamic equilibrium, i.e.,
\begin{equation}
\label{eq22}
\hat {\tilde \rho} _{{R_j}}^{th} = \exp ( - {\hat H_{{R_j}}}/{T_j})/tr(\exp ( - {\hat H_{{R_j}}}/{T_j}))
\end{equation}
Taking trace over reservoir’s variables we obtain the following equations 
\begin{equation}
\label{eq23}
{\hat {\tilde \rho} _{S1}}(t) = 0
\end{equation}

\begin{equation}
\label{eq24}
\begin{array}{l}
{{\hat {\tilde \rho} }_{S2}}({t_0} + \Delta t) = \\
 - \int\limits_{{t_0}}^{{t_0} + \Delta t} {d{t_1}} \int\limits_{{t_0}}^{{t_1}} {d{t_2}} \\ \left( {\begin{array}{*{20}{c}}
\begin{array}{l}
{{\hat {\tilde \rho} }_{S0}}({t_0}){{\hat {\tilde S}}_1}({t_2}){{\hat {\tilde S}}_1}({t_1})T{R_1}({t_2} - {t_1}) - {{\hat {\tilde S}}_1}({t_2}){{\hat {\tilde \rho} }_{S0}}({t_0}){{\hat {\tilde S}}_1}({t_1})T{R_1}( - ({t_2} - {t_1})) - \\
{{\hat {\tilde S}}_1}({t_1}){{\hat {\tilde \rho} }_{S0}}({t_0}){{\hat {\tilde S}}_1}({t_2})T{R_1}({t_2} - {t_1}) + {{\hat {\tilde S}}_1}({t_1}){{\hat {\tilde S}}_1}({t_2}){{\hat {\tilde \rho} }_{S0}}({t_0})T{R_1}( - ({t_2} - {t_1})) + 
\end{array}\\
\begin{array}{l}
{{\hat {\tilde \rho} }_{S0}}({t_0}){{\hat {\tilde S}}_2}({t_2}){{\hat {\tilde S}}_2}({t_1})T{R_2}({t_2} - {t_1}) - {{\hat {\tilde S}}_2}({t_2}){{\hat {\tilde \rho} }_{S0}}({t_0}){{\hat {\tilde S}}_2}({t_1})T{R_2}( - ({t_2} - {t_1})) - \\
{{\hat {\tilde S}}_2}({t_1}){{\hat {\tilde \rho} }_{S0}}({t_0}){{\hat {\tilde S}}_2}({t_2})T{R_2}({t_2} - {t_1}) + {{\hat {\tilde S}}_2}({t_1}){{\hat {\tilde S}}_2}({t_2}){{\hat {\tilde \rho} }_{S0}}({t_0})T{R_2}( - ({t_2} - {t_1}))
\end{array}
\end{array}} \right)
\end{array}
\end{equation}
where $T{R_{1,2}}({t_2} - {t_1}) = Tr(\hat {\tilde \rho} _{{R_1}}^{th}\hat {\tilde \rho} _{{R_2}}^{th}{\hat {\tilde R}_{1,2}}({t_2}){\hat {\tilde R}_{1,2}}({t_1}))$, $T{R_{1,2}}( - ({t_2} - {t_1})) = Tr(\hat {\tilde \rho} _{{R_1}}^{th}\hat {\tilde \rho} _{{R_2}}^{th}{\hat {\tilde R}_{1,2}}({t_1}){\hat {\tilde R}_{1,2}}({t_2}))$ are reservoir correlation function.

To calculate operators ${\hat {\tilde S}_{1,2}}(t)$ we use the Baker–Campbell–Hausdorff formula,

\begin{equation}
\label{eq25}
{e^{\hat A}}\hat B{e^{ - \hat A}} = \hat A + \frac{1}{{1!}}[\hat A,\hat B] + \frac{1}{{2!}}[\hat A,[\hat A,\hat B]] + ...\;
\end{equation}
and diagonal representation of the system Hamiltonian,

\begin{equation}
\label{eq26}
\begin{array}{*{20}{c}}
{{{\hat H}_S} = (\omega_0  + \Omega ){{\hat b}^\dag }\hat b + (\omega_0  - \Omega ){{\hat c}^\dag }\hat c,}\\
{\hat b = \frac{{{{\hat a}_1} + {{\hat a}_2}}}{{\sqrt 2 }},\quad \hat c = \frac{{{{\hat a}_1} - {{\hat a}_2}}}{{\sqrt 2 }},\quad {{\hat b}^\dag } = \frac{{\hat a_1^\dag  + \hat a_2^\dag }}{{\sqrt 2 }},\quad {{\hat c}^\dag } = \frac{{\hat a_1^\dag  - \hat a_2^\dag }}{{\sqrt 2 }}}
\end{array}
\end{equation}

Using commutation relations $[{\hat H_S},\hat b] =  - \,\left( {\omega_0  + \Omega \,} \right)\hat b$, $[{\hat H_S},{\hat b^\dag }] = \left( {\omega_0  + \Omega \,} \right){\hat b^\dag }$, $[{\hat H_S},\hat c] =  - \,\left( {\omega_0  - \Omega \,} \right)\hat c$, $[{\hat H_S},{\hat c^\dag }] = \left( {\omega_0  - \Omega \,} \right){\hat c^\dag }$ and Eq.~\eqref{eq25}, we get 

\begin{equation}
\label{eq27}
\begin{array}{l}
{\hat {\tilde S}_1}(t) =\\ \exp (i{\hat H_S}t)\,({\hat a_1} + \hat a_1^\dag )\exp ( - i{\hat H_S}t) = \frac{1}{{\sqrt 2 }}\left( {b{e^{ - i(\omega_0  + \Omega )t}} + c{e^{ - i(\omega_0  - \Omega )t}} + {b^\dag }{e^{i(\omega_0  + \Omega )t}} + {c^\dag }{e^{i(\omega_0  - \Omega )t}}} \right)
\end{array}
\end{equation}

\begin{equation}
\label{eq28}
\begin{array}{l}
{\hat {\tilde S}_2}(t) =\\ \exp (i{\hat H_S}t)\,({\hat a_2} + \hat a_2^\dag )\exp ( - i{\hat H_S}t) = \frac{1}{{\sqrt 2 }}\left( {b{e^{ - i(\omega_0  + \Omega )t}} - c{e^{ - i(\omega_0  - \Omega )t}} + {b^\dag }{e^{i(\omega_0  + \Omega )t}} - {c^\dag }{e^{i(\omega_0  - \Omega )t}}} \right)
\end{array}
\end{equation}
Now we can substitute expressions for ${\hat {\tilde S}_{1,2}}(t)$ into the equation for ${\hat {\tilde \rho} _{S2}}(t)$. We consider the case $\Omega  \ll \omega_0$ and $\Omega \,\Delta t \ll 1 \ll \omega_0 \,\Delta t$. This enables to estimate the terms that appear from the products like ${\hat {\tilde S}_1}({t_2}){\hat {\tilde S}_1}({t_1})$. The terms proportional to $\hat b\hat b$, ${\hat b^\dag }{\hat b^\dag }$, $\hat c\hat c$, ${\hat c^\dag }{\hat c^\dag }$,  $bc$ and ${\hat b^\dag }{\hat c^\dag }$ after averaging give zero. For instance:

\begin{equation}
\label{eq29}
\begin{array}{l}
 - \int\limits_{{t_0}}^{{t_0} + \Delta t} {d{t_1}} \int\limits_{{t_0}}^{{t_1}} {d{t_2}} {{\hat {\tilde \rho} }_{S0}}({t_0})\hat b{e^{ - i(\omega_0  + \Omega ){t_2}}}\hat c{e^{ - i(\omega_0  - \Omega ){t_1}}}T{R_1}({t_2} - {t_1}) \sim \\
 - \int\limits_{{t_0}}^{{t_0} + \Delta t} {d{t_1}} \int\limits_{{t_0}}^{{t_1}} {d{t_2}} {e^{ - i(\omega_0  + \Omega ){t_2} - i(\omega_0  - \Omega ){t_1}}}T{R_1}({t_2} - {t_1}) \sim \\ \int\limits_{{t_0}}^{{t_0} + \Delta t} {d{t_1}} \int\limits_{{t_0}}^{{t_1}} {d{t_2}} {e^{ - i\omega_0 ({t_2} + {t_1}) - i\Omega ({t_2} - {t_1})}}T{R_1}({t_2} - {t_1}) = \\
 = \left[ {\begin{array}{*{20}{c}}
{\tau  = {t_2} - {t_1}}\\
{d\tau  = d{t_2}}
\end{array}} \right] = \\ \int\limits_{{t_0}}^{{t_0} + \Delta t} {d{t_1}} {e^{ - i2\omega_0 {t_1}}}\int\limits_{{t_0} - {t_1}}^0 {d\tau } {e^{ - i\omega_0 \tau - i\Omega \tau } }T{R_1}(\tau ) \approx \int\limits_{{t_0}}^{{t_0} + \Delta t} {d{t_1}} {e^{ - i2\omega_0 {t_1}}}{G_{1 - }}(\omega_0  + \Omega ) \approx 0,
\end{array}
\end{equation}

where ${G_{1 - }}(\omega )$ is one-side Fourier transform of reservoir correlation function,

\begin{equation}
\label{eq30}
{G_{1 - }}(\omega ) = \int\limits_{ - \infty }^0 {d\tau } {e^{ - i\omega \tau }}T{R_1}(\tau )\,\,\,\,\,\,\,\,\,\,\,\,\,\,\,\,\,\,\,{G_{1 + }}(\omega ) = \int\limits_0^{ + \infty } {d\tau } {e^{ - i\omega \tau }}T{R_1}(\tau )
\end{equation}

\begin{equation}
\label{eq31}
{G_1}(\omega ) = \int\limits_{ - \infty }^{ + \infty } {d\tau } {e^{ - i\omega \tau }}T{R_1}(\tau ) = {G_{1 - }}(\omega ) + {G_{1 + }}(\omega )
\end{equation}
Other terms give one-side Fourier transforms multiplied by slowly oscillating exponents. For example,

\begin{equation}
\label{eq32}
\begin{array}{l}
\int\limits_{{t_0}}^{{t_0} + \Delta t} {d{t_1}} \int\limits_{{t_0}}^{{t_1}} {d{t_2}} {{\hat {\tilde \rho} }_{S0}}({t_0}){{\hat {\tilde S}}_1}({t_2}){{\hat {\tilde S}}_1}({t_1})T{R_1}({t_2} - {t_1}) = \\
\frac{1}{2}{{\hat {\tilde \rho} }_{S0}}({t_0})\int\limits_{{t_0}}^{{t_0} + \Delta t} {d{t_1}} \int\limits_{{t_0}}^{{t_1}} {d{t_2}} \left( {\hat b{{\hat b}^\dag }{e^{ - i(\omega_0  + \Omega )({t_2} - {t_1})}} + {{\hat b}^\dag }\hat b{e^{i(\omega_0  + \Omega )({t_2} - {t_1})}}} \right.\\
 + \hat b{{\hat c}^\dag }{e^{ - i(\omega_0  + \Omega ){t_2} + i(\omega_0  - \Omega ){t_1}}} + {{\hat b}^\dag }\hat c{e^{i(\omega_0  + \Omega ){t_2} - i(\omega_0  - \Omega ){t_1}}} + \hat c{{\hat b}^\dag }{e^{ - i(\omega_0  - \Omega ){t_2} + i(\omega_0  + \Omega ){t_1}}} +\\ {{\hat c}^\dag }\hat b{e^{i(\omega_0  - \Omega ){t_2} - i(\omega_0  + \Omega ){t_1}}} + \left. {\hat c{{\hat c}^\dag }{e^{ - i(\omega_0  - \Omega )({t_2} - {t_1})}} + {{\hat c}^\dag }\hat c{e^{i(\omega_0  - \Omega )({t_2} - {t_1})}}} \right)T{R_1}({t_2} - {t_1}) \approx \\
\frac{1}{2}{{\hat {\tilde \rho} }_{S0}}({t_0})\int\limits_{{t_0}}^{{t_0} + \Delta t} {d{t_1}} \int\limits_{ - \infty }^0 {d\tau } \left( {\hat b{{\hat b}^\dag }{e^{ - i(\omega_0  + \Omega )\tau }} + {{\hat b}^\dag }\hat b{e^{i(\omega_0  + \Omega )\tau }}} \right.
 + \hat b{{\hat c}^\dag }{e^{ - i(\omega_0  + \Omega )\tau  - i2\Omega {t_1}}} +\\ {{\hat b}^\dag }\hat c{e^{i(\omega_0  + \Omega )\tau  + i2\Omega {t_1}}} + \hat c{{\hat b}^\dag }{e^{ - i(\omega_0  - \Omega )\tau  + i2\Omega {t_1}}} + {{\hat c}^\dag }\hat b{e^{i(\omega_0  - \Omega )\tau  + i2\Omega {t_1}}}\\
 + \left. {\hat c{{\hat c}^\dag }{e^{ - i(\omega_0  - \Omega )\tau }} + {{\hat c}^\dag }\hat c{e^{i(\omega_0  - \Omega )\tau }}} \right)T{R_1}(\tau ) =\\ \frac{1}{2}{{\hat {\tilde \rho} }_{S0}}({t_0})\int\limits_{{t_0}}^{{t_0} + \Delta t} {d{t_1}\left( {\hat b{{\hat b}^\dag }{G_{1 - }}(\omega_0  + \Omega ) + {{\hat b}^\dag }\hat b{G_{1 - }}( - (\omega_0  + \Omega )} \right.)} 
 + \hat b{{\hat c}^\dag }{G_{1 - }}(\omega_0  + \Omega ){e^{ - i2\Omega {t_1}}} +\\ {{\hat b}^\dag }\hat c{G_{1 - }}( - (\omega_0  + \Omega )){e^{i2\Omega {t_1}}} + \hat c{{\hat b}^\dag }{G_{1 - }}(\omega_0  - \Omega ){e^{i2\Omega {t_1}}} + {{\hat c}^\dag }\hat b{G_{1 - }}( - (\omega_0  - \Omega )){e^{i2\Omega {t_1}}}\\
 + \left. {\hat c{{\hat c}^\dag }{G_{1 - }}(\omega_0  - \Omega ) + {{\hat c}^\dag }\hat c{G_{1 - }}( - (\omega_0  - \Omega )){e^{i(\omega_0  - \Omega )\tau }}} \right)T{R_1}(\tau )
\end{array}
\end{equation}
We average the equation over the time $1/\Omega  \gg \Delta t \gg 1/\omega_0 $. As a consequence, the terms multiplied by fast-oscillating exponents, $\exp \left( { \pm i\left( {\omega_0  \pm \Omega } \right)t} \right)$, gives zero. Other terms can be integrated straightforwardly. As a result, we obtain (for details, see \cite{ref58}):

\begin{equation}
\label{eq33}
\begin{array}{l}
{\hat {\tilde \rho} _{S2}}({t_0} + \Delta t) - {\hat {\tilde \rho} _{S2}}({t_0}) =\\ \left( { - \left( {\left( I \right) - \left( {II} \right) + \left( {III} \right) - \left( {IV} \right)} \right) - \left( {(1) \to (2),\;\hat c \to  - \hat c,\;{{\hat c}^\dag } \to  - {{\hat c}^\dag }} \right)} \right)\Delta t
\end{array}
\end{equation}
where

\begin{equation}
\label{eq34}
(I) = \frac{1}{2}\left( \begin{array}{l}
{{\hat {\tilde \rho} }_{S0}}\left( {\hat b{{\hat b}^\dag }{G_{1 - }}(\omega_0  + \Omega ) + {{\hat b}^\dag }\hat b{G_{1 - }}( - (\omega_0  + \Omega ))} \right. + \\
\hat b{{\hat c}^\dag }{G_{1 - }}(\omega_0  + \Omega ){e^{ - i2\Omega {t_0}}} + {{\hat b}^\dag }\hat c{G_{1 - }}( - (\omega_0  + \Omega )){e^{i2\Omega {t_0}}} + \\
\hat c{{\hat b}^\dag }{G_{1 - }}(\omega_0  - \Omega ){e^{i2\Omega {t_0}}} + {{\hat c}^\dag }\hat b{G_{1 - }}( - (\omega_0  - \Omega )){e^{ - i2\Omega {t_0}}} + \\
\left. {\hat c{{\hat c}^\dag }{G_{1 - }}(\omega_0  - \Omega ) + {{\hat c}^\dag }\hat c{G_{1 - }}( - (\omega_0  - \Omega ))} \right)
\end{array} \right.
\end{equation}

\begin{equation}
\label{eq35}
(II) = \frac{1}{2}\left( \begin{array}{l}
\left( {\hat b{{\hat {\tilde \rho} }_{S0}}{b^\dag }{G_{1 - }}( - (\omega_0  + \Omega )) + {b^\dag }{{\hat {\tilde \rho} }_{S0}}\hat b{G_{1 - }}(\omega_0  + \Omega )} \right. + \\
\hat b{{\hat {\tilde \rho} }_{S0}}{{\hat c}^\dag }{G_{1 - }}( - (\omega_0  - \Omega )){e^{ - i2\Omega {t_0}}} + {b^\dag }{{\hat {\tilde \rho} }_{S0}}\hat c{G_{1 - }}(\omega_0  - \Omega ){e^{i2\Omega {t_0}}} + \\
\hat c{{\hat {\tilde \rho} }_{S0}}{b^\dag }{G_{1 - }}( - (\omega_0  + \Omega )){e^{i2\Omega {t_0}}} + {{\hat c}^\dag }{{\hat {\tilde \rho} }_{S0}}\hat b{G_{1 - }}(\omega_0  + \Omega ){e^{ - i2\Omega {t_0}}} + \\
\left. {\hat c{{\hat {\tilde \rho} }_{S0}}{{\hat c}^\dag }{G_{1 - }}( - (\omega_0  - \Omega )) + {{\hat c}^\dag }{{\hat {\tilde \rho} }_{S0}}\hat c{G_{1 - }}(\omega_0  - \Omega )} \right)
\end{array} \right.
\end{equation}

\begin{equation}
\label{eq36}
(III) = \frac{1}{2}\left( \begin{array}{l}
\left( {\hat b{{\hat b}^\dag }{G_{1 + }}(\omega_0  + \Omega ) + {{\hat b}^\dag }\hat b{G_{1 + }}( - (\omega_0  + \Omega ))} \right. + \\
\hat b{{\hat c}^\dag }{G_{1 + }}(\omega_0  - \Omega ){e^{ - i2\Omega {t_0}}} + {{\hat b}^\dag }\hat c{G_{1 + }}( - (\omega_0  - \Omega )){e^{i2\Omega {t_0}}} + \\
\hat c{{\hat b}^\dag }{G_{1 + }}(\omega_0  + \Omega ){e^{i2\Omega {t_0}}} + {{\hat c}^\dag }\hat b{G_{1 + }}( - (\omega_0  + \Omega )){e^{ - i2\Omega {t_0}}} + \\
\left. {\hat c{{\hat c}^\dag }{G_{1 + }}(\omega_0  - \Omega ) + {{\hat c}^\dag }\hat c{G_{1 + }}( - (\omega_0  - \Omega ))} \right){{\hat {\tilde \rho} }_{S0}}
\end{array} \right.
\end{equation}

\begin{equation}
\label{eq37}
(IV) = \frac{1}{2}\left( \begin{array}{l}
\left( {\hat b{{\hat {\tilde \rho} }_{S0}}{{\hat b}^\dag }{G_{1 + }}( - (\omega_0  + \Omega )) + {{\hat b}^\dag }{{\hat {\tilde \rho} }_{S0}}\hat b{G_{1 + }}(\omega_0  + \Omega )} \right. + \\
\hat b{{\hat {\tilde \rho} }_{S0}}{{\hat c}^\dag }{G_{1 + }}( - (\omega_0  + \Omega )){e^{ - i2\Omega {t_0}}} + {{\hat b}^\dag }{{\hat {\tilde \rho} }_{S0}}\hat c{G_{1 + }}(\omega_0  + \Omega ){e^{i2\Omega {t_0}}} + \\
\hat c{{\hat {\tilde \rho} }_{S0}}{{\hat b}^\dag }{G_{1 + }}( - (\omega_0  - \Omega )){e^{i2\Omega {t_0}}} + {{\hat c}^\dag }{{\hat {\tilde \rho} }_{S0}}\hat b{G_{1 + }}(\omega_0  - \Omega ){e^{ - i2\Omega {t_0}}} + \\
\left. {\hat c{{\hat {\tilde \rho} }_{S0}}{{\hat c}^\dag }{G_{1 + }}( - (\omega_0  - \Omega )) + {{\hat c}^\dag }{{\hat {\tilde \rho} }_{S0}}\hat c{G_{1 + }}(\omega_0  - \Omega )} \right)
\end{array} \right.
\end{equation}
and $(1) \to (2)$ denotes same terms with index exchange. Further, we divide both parts of Eq.~\eqref{eq33} by $\Delta t$ and replace $\left( {{{\hat {\tilde \rho} }_{S2}}({t_0} + \Delta t) - {{\hat {\tilde \rho} }_{S2}}({t_0})} \right)/\Delta t$  with $\partial {\hat {\tilde \rho} _S}\left( {{t_0}} \right)/\partial t$. Finally, we get

\begin{equation}
\label{eq38}
\begin{array}{l}
\frac{1}{{{\lambda ^2}}}\frac{{\partial {{\hat {\tilde \rho} }_S}}}{{\partial t}} = \\
 + \frac{{{G_1}( - (\omega_0  + \Omega ))}}{4}\left( {2\hat b{{\hat {\tilde \rho} }_S}{{\hat b}^\dag } - {{\hat {\tilde \rho} }_S}{{\hat b}^\dag }\hat b - {{\hat b}^\dag }\hat b{{\hat {\tilde \rho} }_S}} \right) + \frac{{{G_1}(\omega_0  + \Omega )}}{4}\left( {2{{\hat b}^\dag }{{\hat {\tilde \rho} }_S}\hat b - {{\hat {\tilde \rho} }_S}\hat b{{\hat b}^\dag } - \hat b{{\hat b}^\dag }{{\hat {\tilde \rho} }_S}} \right) + \\
 + \frac{{{G_1}( - (\omega_0  - \Omega ))}}{4}\left( {2\hat c{{\hat {\tilde \rho} }_S}{{\hat c}^\dag } - {{\hat {\tilde \rho} }_S}{{\hat c}^\dag }\hat c - {{\hat c}^\dag }\hat c{{\hat {\tilde \rho} }_S}} \right) + \frac{{{G_1}(\omega_0  - \Omega )}}{4}\left( {2{{\hat c}^\dag }{{\hat {\tilde \rho} }_S}\hat c - {{\hat {\tilde \rho} }_S}\hat c{{\hat c}^\dag } - \hat c{{\hat c}^\dag }{{\hat {\tilde \rho} }_S}} \right) + \\
 + \frac{{{G_{1 - }}( - (\omega_0  - \Omega )) + {G_{1 + }}( - (\omega_0  + \Omega ))}}{4}\left( {2\hat b{{\hat {\tilde \rho} }_S}{{\hat c}^\dag } - {{\hat {\tilde \rho} }_S}{{\hat c}^\dag }\hat b - {{\hat c}^\dag }\hat b{{\hat {\tilde \rho} }_S}} \right){e^{ - i2\Omega t}} + \\
 + \frac{{{G_{1 - }}(\omega_0  + \Omega ) + {G_{1 + }}(\omega_0  - \Omega )}}{4}\left( {2{{\hat c}^\dag }{{\hat {\tilde \rho} }_S}\hat b - {{\hat {\tilde \rho} }_S}\hat b{{\hat c}^\dag } - \hat b{{\hat c}^\dag }{{\hat {\tilde \rho} }_S}} \right){e^{ - i2\Omega t}} + \\
 + \frac{{{G_{1 - }}(\omega_0  - \Omega ) + {G_{1 + }}(\omega_0  + \Omega )}}{4}\left( {2{{\hat b}^\dag }{{\hat {\tilde \rho} }_S}\hat c - {{\hat {\tilde \rho} }_S}\hat c{{\hat b}^\dag } - \hat c{{\hat b}^\dag }{{\hat {\tilde \rho} }_S}} \right){e^{i2\Omega t}} + \\
 + \frac{{{G_{1 - }}( - (\omega_0  + \Omega )) + {G_{1 + }}( - (\omega_0  - \Omega ))}}{4}\left( {2\hat c{{\hat {\tilde \rho} }_S}{{\hat b}^\dag } - {{\hat {\tilde \rho} }_S}{{\hat b}^\dag }\hat c - {{\hat b}^\dag }\hat c{{\hat {\tilde \rho} }_S}} \right){e^{i2\Omega t}} - \\
 - \frac{{{G_{1 + }}( - (\omega_0  + \Omega )) - {G_{1 - }}( - (\omega_0  + \Omega )) + {G_{1 + }}(\omega_0  + \Omega ) - {G_{1 - }}(\omega_0  + \Omega )}}{4}\left[ {{{\hat b}^\dag }\hat b,{{\hat {\tilde \rho} }_S}} \right] - \\
 - \frac{{{G_{1 + }}( - (\omega_0  - \Omega )) - {G_{1 - }}( - (\omega_0  - \Omega )) + {G_{1 + }}(\omega_0  - \Omega ) - {G_{1 - }}(\omega_0  - \Omega )}}{4}\left[ {{{\hat c}^\dag }\hat c,{{\hat {\tilde \rho} }_S}} \right] + \\
 + \frac{{{G_{1 - }}( - (\omega_0  - \Omega )) - {G_{1 + }}( - (\omega_0  + \Omega ))}}{4}\left[ {{{\hat c}^\dag }\hat b,{{\hat {\tilde \rho} }_S}} \right]{e^{ - i2\Omega t}} + \frac{{{G_{1 - }}(\omega_0  + \Omega ) - {G_{1 + }}(\omega_0  - \Omega )}}{4}\left[ {\hat b{{\hat c}^\dag },{{\hat {\tilde \rho} }_S}} \right]{e^{ - i2\Omega t}} + \\
 + \frac{{{G_{1 - }}(\omega_0  - \Omega ) - {G_{1 + }}(\omega_0  + \Omega )}}{4}\left[ {\hat c{{\hat b}^\dag },{{\hat {\tilde \rho} }_S}} \right]{e^{i2\Omega t}} + \frac{{{G_{1 - }}( - (\omega_0  + \Omega )) - {G_{1 + }}( - (\omega_0  - \Omega ))}}{4}\left[ {{{\hat b}^\dag }\hat c,{{\hat {\tilde \rho} }_S}} \right]{e^{i2\Omega t}} + \\
 + \left( {(1) \to (2),\;\hat c \to  - \hat c,\;{{\hat c}^\dag } \to  - {{\hat c}^\dag }} \right).
\end{array}
\end{equation}

This approximation is called semisecular \cite{ref75}. The difference from the secular approach and master equation in the Lindblad form is in slowly oscillating terms. These terms are absent in secular approximation where slowly oscillating exponents are averaged.

One can move back from interaction picture to the Schrodinger picture via formulas

\begin{equation}
\label{eq39}
\dot {\hat \rho}  = d\left( {\exp \left( { - i{{\hat H}_S}t} \right)\hat {\tilde \rho} \exp \left( {i{{\hat H}_S}t} \right)} \right)/dt =  - i\left[ {{{\hat H}_S},\hat \rho } \right] + \exp \left( { - i{{\hat H}_S}t} \right)\frac{{d\hat {\tilde \rho} }}{{dt}}\exp \left( {i{{\hat H}_S}t} \right)
\end{equation}

\begin{equation}
\label{eq40}
\begin{array}{*{20}{c}}
{\exp \left( { - i{{\hat H}_S}t} \right)\left( {2\hat X(t)\hat {\tilde \rho} (t)\hat Y(t) - \hat {\tilde \rho} (t)\hat Y(t)\hat X(t) - \hat Y(t)\hat X(t)\hat {\tilde \rho} (t)} \right)\exp \left( {i{{\hat H}_S}t} \right) = }\\
{ = \left( {2\hat {\tilde X}( - t)\hat \rho (t)\hat {\tilde Y}( - t) - \hat \rho (t)\hat {\tilde Y}( - t)\hat {\tilde X}( - t) - \hat {\tilde Y}( - t)\hat {\tilde X}( - t)\hat \rho (t)} \right)}
\end{array}
\end{equation}
Straightforward calculations give us the following master equation

\begin{equation}
\label{eq41}
\begin{array}{l}
\frac{1}{{{\lambda ^2}}}\frac{{\partial {{\hat \rho }_S}}}{{\partial t}} =  - \frac{i}{{{\lambda ^2}}}[{{\hat H}_S},{{\hat \rho }_S}] + \\
 + \frac{{{G_1}( - (\omega_0  + \Omega ))}}{4}\left( {2\hat b{{\hat \rho }_S}{{\hat b}^\dag } - {{\hat \rho }_S}{{\hat b}^\dag }\hat b - {{\hat b}^\dag }\hat b{{\hat \rho }_S}} \right) + \frac{{{G_1}(\omega_0  + \Omega )}}{4}\left( {2{{\hat b}^\dag }{{\hat \rho }_S}\hat b - {{\hat \rho }_S}\hat b{{\hat b}^\dag } - \hat b{{\hat b}^\dag }{{\hat \rho }_S}} \right) + \\
 + \frac{{{G_1}( - (\omega_0  - \Omega ))}}{4}\left( {2\hat c{{\hat \rho }_S}{{\hat c}^\dag } - {{\hat \rho }_S}{{\hat c}^\dag }\hat c - {{\hat c}^\dag }\hat c{{\hat \tilde \rho }_S}} \right) + \frac{{{G_1}(\omega_0  - \Omega )}}{4}\left( {2{{\hat c}^\dag }{{\hat \rho }_S}\hat c - {{\hat \rho }_S}\hat c{{\hat c}^\dag } - \hat c{{\hat c}^\dag }{{\hat \rho }_S}} \right) + \\
 + \frac{{{G_{1 - }}( - (\omega_0  - \Omega )) + {G_{1 + }}( - (\omega_0  + \Omega ))}}{4}\left( {2\hat b{{\hat \rho }_S}{{\hat c}^\dag } - {{\hat \rho }_S}{{\hat c}^\dag }\hat b - {{\hat c}^\dag }\hat b{{\hat \rho }_S}} \right) + \\
 + \frac{{{G_{1 - }}(\omega_0  + \Omega ) + {G_{1 + }}(\omega_0  - \Omega )}}{4}\left( {2{{\hat c}^\dag }{{\hat \rho }_S}\hat b - {{\hat \rho }_S}\hat b{{\hat c}^\dag } - \hat b{{\hat c}^\dag }{{\hat \rho }_S}} \right) + \\
 + \frac{{{G_{1 - }}(\omega_0  - \Omega ) + {G_{1 + }}(\omega_0  + \Omega )}}{4}\left( {2{{\hat b}^\dag }{{\hat \rho }_S}\hat c - {{\hat \rho }_S}\hat c{{\hat b}^\dag } - \hat c{{\hat b}^\dag }{{\hat \rho }_S}} \right) + \\
 + \frac{{{G_{1 - }}( - (\omega_0  + \Omega )) + {G_{1 + }}( - (\omega_0  - \Omega ))}}{4}\left( {2\hat c{{\hat \rho }_S}{{\hat b}^\dag } - {{\hat \rho }_S}{{\hat b}^\dag }\hat c - {{\hat b}^\dag }\hat c{{\hat \rho }_S}} \right) - \\
 - \frac{{{G_{1 + }}( - (\omega_0  + \Omega )) - {G_{1 - }}( - (\omega_0  + \Omega )) + {G_{1 + }}(\omega_0  + \Omega ) - {G_{1 - }}(\omega_0  + \Omega )}}{4}\left[ {{{\hat b}^\dag }\hat b,{{\hat \rho }_S}} \right] - \\
 - \frac{{{G_{1 + }}( - (\omega_0  - \Omega )) - {G_{1 - }}( - (\omega_0  - \Omega )) + {G_{1 + }}(\omega_0  - \Omega ) - {G_{1 - }}(\omega_0  - \Omega )}}{4}\left[ {{{\hat c}^\dag }\hat c,{{\hat \rho }_S}} \right] + \\
 + \frac{{{G_{1 - }}( - (\omega_0  - \Omega )) - {G_{1 + }}( - (\omega_0  + \Omega ))}}{4}\left[ {{{\hat c}^\dag }\hat b,{{\hat \rho }_S}} \right] + \frac{{{G_{1 - }}(\omega_0  + \Omega ) - {G_{1 + }}(\omega_0  - \Omega )}}{4}\left[ {\hat b{{\hat c}^\dag },{{\hat \rho }_S}} \right] + \\
 + \frac{{{G_{1 - }}(\omega_0  - \Omega ) - {G_{1 + }}(\omega_0  + \Omega )}}{4}\left[ {\hat c{{\hat b}^\dag },{{\hat \rho }_S}} \right] + \frac{{{G_{1 - }}( - (\omega_0  + \Omega )) - {G_{1 + }}( - (\omega_0  - \Omega ))}}{4}\left[ {{{\hat b}^\dag }\hat c,{{\hat \rho }_S}} \right] + \\
 + \left( {(1) \to (2),\;\hat c \to  - \hat c,\;{{\hat c}^\dag } \to  - {{\hat c}^\dag }} \right)
\end{array}
\end{equation}

Now let’s compute $d\hat a/dt = Tr\left( {\dot {\hat \rho} \,\hat a} \right)$. For this purpose we use the following identity

\begin{equation}
\label{eq42}
\begin{array}{l}
Tr(2\hat X{{\hat \rho }_S}\hat Y\hat a - {{\hat \rho }_S}\hat Y\hat X\hat a - \hat Y\hat X{{\hat \rho }_S}\hat a) = 2Tr({{\hat \rho }_S}\hat Y\hat a\hat X) - Tr({{\hat \rho }_S}\hat Y\hat X\hat a) - Tr({{\hat \rho }_S}\hat a\hat Y\hat X) = \\
 = 2\left\langle {\hat Y\hat a\hat X} \right\rangle  - \left\langle {\hat Y\hat X\hat a} \right\rangle  - \left\langle {\hat a\hat Y\hat X} \right\rangle  = \left\langle {\hat Y\left[ {\hat a,\hat X} \right]} \right\rangle  + \left\langle {\left[ {\hat Y,\hat a} \right]\hat X} \right\rangle 
\end{array}
\end{equation}
Applying this formula, we get

\begin{equation}
\label{eq43}
\begin{array}{l}
2Tr\left( {{{\hat \rho }_S}{{\hat b}^\dag }\hat b\hat b} \right) - Tr\left( {{{\hat \rho }_S}{{\hat b}^\dag }\hat b\hat b} \right) - Tr\left( {{{\hat \rho }_S}\hat b{{\hat b}^\dag }\hat b} \right) = \left\langle {{{\hat b}^\dag }\left[ {\hat b,\hat b} \right]} \right\rangle  + \left\langle {\left[ {{{\hat b}^\dag },\hat b} \right]\hat b} \right\rangle  =  - \left\langle {\hat b} \right\rangle ,\\
2Tr\left( {{{\hat \rho }_S}\hat b\hat b{{\hat b}^\dag }} \right) - Tr\left( {{{\hat \rho }_S}\hat b{{\hat b}^\dag }\hat b} \right) - Tr\left( {{{\hat \rho }_S}\hat b\hat b{{\hat b}^\dag }} \right) = \left\langle {\hat b\left[ {\hat b,{{\hat b}^\dag }} \right]} \right\rangle  + \left\langle {\left[ {\hat b,\hat b} \right]{{\hat b}^\dag }} \right\rangle  = \left\langle {\hat b} \right\rangle ,\\
2Tr\left( {{{\hat \rho }_S}{{\hat c}^\dag }\hat b\hat c} \right) - Tr\left( {{{\hat \rho }_S}{{\hat c}^\dag }\hat c\hat b} \right) - Tr\left( {{{\hat \rho }_S}\hat b{{\hat c}^\dag }\hat c} \right) = \left\langle {{{\hat c}^\dag }\left[ {\hat b,\hat c} \right]} \right\rangle  + \left\langle {\left[ {{{\hat c}^\dag },\hat b} \right]\hat c} \right\rangle  = 0,\\
2Tr\left( {{{\hat \rho }_S}\hat c\hat b{{\hat c}^\dag }} \right) - Tr\left( {{{\hat \rho }_S}\hat c{{\hat c}^\dag }\hat b} \right) - Tr\left( {{{\hat \rho }_S}\hat b\hat c{{\hat c}^\dag }} \right) = \left\langle {\hat c\left[ {\hat b,{{\hat c}^\dag }} \right]} \right\rangle  + \left\langle {\left[ {\hat c,\hat b} \right]{{\hat c}^\dag }} \right\rangle  = 0,\\
2Tr\left( {{{\hat \rho }_S}{{\hat c}^\dag }\hat b\hat b} \right) - Tr\left( {{{\hat \rho }_S}{{\hat c}^\dag }\hat b\hat b} \right) - Tr\left( {{{\hat \rho }_S}\hat b{{\hat c}^\dag }\hat b} \right) = \left\langle {{{\hat c}^\dag }\left[ {\hat b,\hat b} \right]} \right\rangle  + \left\langle {\left[ {{{\hat c}^\dag },\hat b} \right]\hat b} \right\rangle  = 0,\\
2Tr\left( {{{\hat \rho }_S}\hat b\hat b{{\hat c}^\dag }} \right) - Tr\left( {{{\hat \rho }_S}\hat b{{\hat c}^\dag }\hat b} \right) - Tr\left( {{{\hat \rho }_S}\hat b\hat b{{\hat c}^\dag }} \right) = \left\langle {\hat b\left[ {\hat b,{{\hat c}^\dag }} \right]} \right\rangle  + \left\langle {\left[ {\hat b,\hat b} \right]{{\hat c}^\dag }} \right\rangle  = 0,\\
2Tr\left( {{{\hat \rho }_S}\hat c\hat b{{\hat b}^\dag }} \right) - Tr\left( {{{\hat \rho }_S}\hat c{{\hat b}^\dag }\hat b} \right) - Tr\left( {{{\hat \rho }_S}\hat b\hat c{{\hat b}^\dag }} \right) = \left\langle {\hat c\left[ {\hat b,{{\hat b}^\dag }} \right]} \right\rangle  + \left\langle {\left[ {\hat c,\hat b} \right]{{\hat b}^\dag }} \right\rangle  = \left\langle {\hat c} \right\rangle ,\\
2Tr\left( {{{\hat \rho }_S}{{\hat b}^\dag }\hat b\hat c} \right) - Tr\left( {{{\hat \rho }_S}{{\hat b}^\dag }\hat c\hat b} \right) - Tr\left( {{{\hat \rho }_S}\hat b{{\hat b}^\dag }\hat c} \right) = \left\langle {{{\hat b}^\dag }\left[ {\hat b,\hat c} \right]} \right\rangle  + \left\langle {\left[ {{{\hat b}^\dag },\hat b} \right]\hat c} \right\rangle  =  - \left\langle {\hat c} \right\rangle ,\\
\begin{array}{*{20}{c}}
{Tr\left( {\left[ {{{\hat c}^\dag }\hat b,{{\hat \rho }_0}} \right]\hat b} \right) = 0,}&{Tr\left( {\left[ {\hat c{{\hat b}^\dag },{{\hat \rho }_0}} \right]\hat b} \right) = \left\langle {\hat c} \right\rangle ,}\\
{Tr\left( {\left[ {{{\hat c}^\dag }\hat b,{{\hat \rho }_S}} \right]\hat c} \right) = \left\langle {\hat b} \right\rangle ,}&{Tr\left( {\left[ {\hat c{{\hat b}^\dag },{{\hat \rho }_S}} \right]\hat c} \right) = 0.}
\end{array}
\end{array}
\end{equation}
Combining obtained terms and neglecting last four terms in Eq.~\eqref{eq41} (they just result in small changing of eigenmode frequencies), we arrive at the following equations for average amplitudes of eigenmodes

\begin{equation}
\label{eq44}
\begin{array}{l}
\frac{{d\left\langle {\hat b} \right\rangle }}{{dt}} =  - i(\omega_0  + \Omega )\left\langle {\hat b} \right\rangle  + {\lambda ^2}\left( {A\left\langle {\hat b} \right\rangle  + (B + C)\left\langle {\hat c} \right\rangle } \right)\\
\frac{{d\left\langle {\hat c} \right\rangle }}{{dt}} =  - i(\omega_0  - \Omega )\left\langle {\hat c} \right\rangle  + {\lambda ^2}\left( {\tilde A\left\langle {\hat c} \right\rangle  + (\tilde B + \tilde C)\left\langle {\hat b} \right\rangle } \right)
\end{array}
\end{equation}
where

\begin{equation}
\label{eq45}
\begin{array}{l}
A =  - \frac{{{G_1}( - (\omega_0  + \Omega )) + {G_2}( - (\omega_0  + \Omega ))}}{4} + \frac{{{G_1}(\omega_0  + \Omega ) + {G_2}(\omega_0  + \Omega )}}{4},\\
\tilde A =  - \frac{{{G_1}( - (\omega_0  - \Omega )) + {G_2}( - (\omega_0  - \Omega ))}}{4} + \frac{{{G_1}(\omega_0  - \Omega ) + {G_2}(\omega_0  - \Omega )}}{4},\\
B + C = \frac{{{G_{1 - }}(\omega_0  - \Omega ) - {G_{1 + }}( - (\omega_0  - \Omega )) - {G_{2 - }}(\omega_0  - \Omega ) + {G_{2 + }}( - (\omega_0  - \Omega ))}}{2},\\
\tilde B + \tilde C = \frac{{{G_{1 - }}(\omega_0  + \Omega ) - {G_{1 + }}( - (\omega_0  + \Omega )) - {G_{2 - }}(\omega_0  + \Omega ) + {G_{2 + }}( - (\omega_0  + \Omega ))}}{2},\\
B = \frac{{{G_{1 - }}(\omega_0  - \Omega ) + {G_{1 + }}(\omega_0  + \Omega )}}{4} - \frac{{{G_{2 - }}(\omega_0  - \Omega ) + {G_{2 + }}(\omega_0  + \Omega )}}{4} - \\
\quad \;\; - \frac{{{G_{1 - }}( - (\omega_0  + \Omega )) + {G_{1 + }}( - (\omega_0  - \Omega ))}}{4} + \frac{{{G_{2 - }}( - (\omega_0  + \Omega )) + {G_{2 + }}( - (\omega_0  - \Omega ))}}{4},\\
\tilde B = \frac{{{G_{1 - }}(\omega_0  + \Omega ) + {G_{1 + }}(\omega_0  - \Omega )}}{4} - \frac{{{G_{2 - }}(\omega_0  + \Omega ) + {G_{2 + }}(\omega_0  - \Omega )}}{4} - \\
\quad \;\; - \frac{{{G_{1 - }}( - (\omega_0  - \Omega )) + {G_{1 + }}( - (\omega_0 + \Omega ))}}{4} + \frac{{{G_{2 - }}( - (\omega_0  - \Omega )) + {G_{2 + }}( - (\omega_0  + \Omega ))}}{4},\\
C = \frac{{{G_{1 - }}( - (\omega_0  + \Omega )) - {G_{1 + }}( - (\omega_0  - \Omega )) + {G_{1 - }}(\omega_0  - \Omega ) - {G_{1 + }}(\omega_0  + \Omega )}}{4} - \\
\quad \;\; - \frac{{{G_{2 - }}( - (\omega_0  + \Omega )) - {G_{2 + }}( - (\omega_0  - \Omega )) + {G_{2 - }}(\omega_0  - \Omega ) - {G_{2 + }}(\omega_0  + \Omega )}}{4},\\
\tilde C = \frac{{{G_{1 - }}( - (\omega_0  - \Omega )) - {G_{1 + }}( - (\omega_0  + \Omega )) + {G_{1 - }}(\omega_0  + \Omega ) - {G_{1 + }}(\omega_0  - \Omega )}}{4} - \\
\quad \;\; - \frac{{{G_{2 - }}( - (\omega_0  - \Omega )) - {G_{2 + }}( - (\omega_0  + \Omega )) + {G_{2 - }}(\omega_0  + \Omega ) - {G_{2 + }}(\omega_0  - \Omega )}}{4}
\end{array}
\end{equation}
The equation for average amplitudes of eigenmodes ~\eqref{eq44} can be rewritten in terms of average amplitudes of oscillators via the linear transformation:

\begin{equation}
\label{eq46}
\begin{array}{l}
d\left\langle {{{\hat a}_1}} \right\rangle /dt =  - i\omega_0 \left\langle {{{\hat a}_1}} \right\rangle  - i\Omega \left\langle {{{\hat a}_2}} \right\rangle  + {\lambda ^2}\frac{{A + \tilde A + B + C + \tilde B + \tilde C}}{2}\left\langle {{{\hat a}_1}} \right\rangle  + {\lambda ^2}\frac{{(A - \tilde A) - (B + C - \tilde B - \tilde C)}}{2}\left\langle {{{\hat a}_2}} \right\rangle \\
d\left\langle {{{\hat a}_2}} \right\rangle /dt =  - i\omega_0 \left\langle {{{\hat a}_2}} \right\rangle  - i\Omega \left\langle {{{\hat a}_1}} \right\rangle  + {\lambda ^2}\frac{{A + \tilde A - (B + C + \tilde B + \tilde C)}}{2}\left\langle {{{\hat a}_2}} \right\rangle  + {\lambda ^2}\frac{{(A - \tilde A) + (B + C - \tilde B - \tilde C)}}{2}\left\langle {{{\hat a}_1}} \right\rangle 
\end{array}
\end{equation}
Calculating ${G_{1,2 \pm }}(\omega )$ using Eqs.~\eqref{eq22} and~\eqref{eq30}, we get following equations for oscillators amplitudes
\begin{equation}
\label{eq47}
\frac{d}{{dt}}\left( {\begin{array}{*{20}{c}}
{\left\langle {{{\hat a}_1}} \right\rangle }\\
{\left\langle {{{\hat a}_2}} \right\rangle }
\end{array}} \right) = \left( {\begin{array}{*{20}{c}}
{ - i\omega_0  - {\lambda ^2}\frac{{{\gamma _1}(\omega_0  + \Omega ) + {\gamma _1}(\omega_0  - \Omega )}}{2}}&{ - i\Omega  - {\lambda ^2}\frac{{{\gamma _1}(\omega_0  + \Omega ) - {\gamma _1}(\omega_0  - \Omega )}}{2}}\\
{ - i\Omega  - {\lambda ^2}\frac{{{\gamma _2}(\omega_0  + \Omega ) - {\gamma _2}(\omega_0  - \Omega )}}{2}}&{ - i\omega_0  - {\lambda ^2}\frac{{{\gamma _2}(\omega_0  + \Omega ) + {\gamma _2}(\omega_0  - \Omega )}}{2}}
\end{array}} \right)\left( {\begin{array}{*{20}{c}}
{\left\langle {{{\hat a}_1}} \right\rangle }\\
{\left\langle {{{\hat a}_2}} \right\rangle }
\end{array}} \right)
\end{equation}
Here ${\gamma _{1,2}}(\omega ) = \pi {\left( {\gamma _k^{(1),(2)}} \right)^2}{\rho _{1,2}}(\omega )$.

\section{Classical equations for two coupled oscillators interacting with environment}
We consider a system of two identical oscillators coupling with each other. Each oscillator interacts with its own reservoir. This system is described by the following system equations

\begin{equation}
\label{eq48}
{\ddot x_1} + \omega _0^2{x_1} = \mathop \sum \limits_\omega  \rho _\omega ^{(1)}g_\omega ^{(1)}y_\omega ^{(1)} - {\kappa ^2}{x_2}
\end{equation}

\begin{equation}
\label{eq49}
{\ddot x_2} + \omega _0^2{x_2} = \mathop \sum \limits_\omega  \rho _\omega ^{(2)}g_\omega ^{(2)}y_\omega ^{(2)} - {\kappa ^2}{x_1}
\end{equation}

\begin{equation}
\label{eq50}
\ddot y_\omega ^{(1)} + {\omega ^2}y_\omega ^{(1)} = g_\omega ^{(1)}{x_1}
\end{equation}

\begin{equation}
\label{eq51}
\ddot y_\omega ^{(2)} + {\omega ^2}y_\omega ^{(2)} = g_\omega ^{(2)}{x_2}
\end{equation}
Here ${x_1}$ and ${x_2}$ are coordinates of first and second oscillators; $y_k^{\left( 1 \right)}$ and $y_k^{\left( 2 \right)}$ are coordinates of kth oscillator in first and second reservoirs, respectively. $\kappa $ is a coupling constant between the oscillators. $g_\omega ^{(1)}$ and $g_\omega ^{(2)}$ are, respectively, the coupling strength between the first and second oscillators with the modes of first and second reservoirs with a frequency $\omega$. $\rho _\omega ^{(1)}$ and $\rho _\omega ^{(2)}$ are, respectively, the number of modes in the first and second reservoirs with a frequency $\omega$.

We exclude the reservoir variables, $y_k^{\left( 1 \right)}$ and $y_k^{\left( 2 \right)}$, in the Born approximation \cite{ref71,ref72}. In this approach, we obtain a closed system of equations for the coordinates of two oscillators. The interaction of oscillators with the reservoirs leads to the appearance of relaxation terms in the equations. It is usually assumed that each oscillator has its own relaxation rate, which does not depend on the amplitude of the other oscillator. However, this assumption is not correct. To exclude the degrees of freedom of the reservoirs it is necessary to consider the interaction of reservoirs with the eigenstates of the system of oscillators \cite{ref52}-\cite{ref54}, and not with the individual oscillators. As a result, the cross-relaxation terms appear in the equations for oscillator coordinates.

The eigenstates of the system of two coupled oscillators are symmetric and anti-symmetric modes, i.e., ${\vec h_{s,a}} = \frac{1}{{\sqrt 2 }}{\left( {1,\,\, \pm 1} \right)^T}$; the eigenfrequencies are $\omega _{s,a}^2 = \omega _0^2 \pm {\kappa ^2}$. Introducing the vector $\vec x(t) = {\left( {{x_1},\,\,{x_2}} \right)^T}$, we can write the system evolution as 
\begin{equation}
\label{eq52}
\vec x(t) = {C_s}(t){\vec h_s}{e^{-i{\omega _s}t}} + {C_a}(t){\vec h_a}{e^{-i{\omega _a}t}} + c.c.
\end{equation}
where ${C_{{\rm{s}}{\rm{,a}}}}$ are complex amplitudes. At this step, we move from real to complex variables and increase the number of unknown variables. We also can introduce symmetric and anti-symmetric modes for the reservoir variables

\begin{equation}
\label{eq53}
y_\omega ^{(s)} = \frac{1}{{\sqrt {g_\omega ^{{{(1)}^2}} + g_\omega ^{{{(2)}^2}}} }}(g_\omega ^{(2)}y_\omega ^{(1)} + g_\omega ^{(1)}y_\omega ^{(2)}) = \mu _\omega ^{(s)}{e^{-i{\omega _s}t}} + \mu _\omega ^{(s)*}{e^{i{\omega _s}t}}
\end{equation}

\begin{equation}
\label{eq54}
y_\omega ^{(a)} = \frac{1}{{\sqrt {g_\omega ^{{{(1)}^2}} + g_\omega ^{{{(2)}^2}}} }}(g_\omega ^{(2)}y_\omega ^{(1)} - g_\omega ^{(1)}y_\omega ^{(2)}) = \mu _\omega ^{(a)}{e^{-i{\omega _a}t}} + \mu _\omega ^{(a)*}{e^{i{\omega _a}t}}
\end{equation}
Following \cite{ref73}, we impose constraints on the first derivatives of complex amplitudes ${C_{s,a}}(t)$, ${\dot C_{s,a}}(t)exp\left( {-i{\omega _{s,a}}t} \right) + {\dot C_{s,a}}^*(t)exp\left( {i{\omega _{s,a}}t} \right) = 0$ and restore the number of independent unknown variables. Using these constraints, we obtain

\begin{equation}
\label{eq55}
{\dot C_s} =\frac{i}{{{\omega _s}}}\sum\limits_\omega  {\left( {g_\omega ^{\left( {ss} \right)}\mu _\omega ^{(s)} + g_\omega ^{(sa)}\mu _\omega ^{(a)}{e^{-i\left( {{\omega _a} - {\omega _s}} \right)t}}} \right)}
\end{equation}

\begin{equation}
\label{eq56}
{\dot C_a} =\frac{i}{{{\omega _a}}}\sum\limits_\omega  {\left( {g_\omega ^{\left( {sa} \right)}\mu _\omega ^{(s)}{e^{i\left( {{\omega _a} - {\omega _s}} \right)t}} + g_\omega ^{(ss)}\mu _\omega ^{(a)}} \right)}
\end{equation}
where $g_\omega ^{\left( {ss} \right)} = \sqrt {g_\omega ^{{{(1)}^2}} + g_\omega ^{{{(2)}^2}}} \frac{{\rho _\omega ^{(1)}g_\omega ^{{{(1)}^2}} + \rho _\omega ^{(2)}g_\omega ^{{{(2)}^2}}}}{{{2^{3/2}}g_\omega ^{(2)}g_\omega ^{(1)}}}$,  $g_\omega ^{\left( {sa} \right)} = \sqrt {g_\omega ^{{{(1)}^2}} + g_\omega ^{{{(2)}^2}}} \frac{{\rho _\omega ^{(1)}g_\omega ^{{{(1)}^2}} - \rho _\omega ^{(2)}g_\omega ^{{{(2)}^2}}}}{{{2^{3/2}}g_\omega ^{(2)}g_\omega ^{(1)}}}$.
Moreover, from Eqs.~\eqref{eq50} and~\eqref{eq51} we obtain

\begin{equation}
\label{eq57}
\dot \mu _\omega ^{(s)} + 2i{\Delta _s}\mu _\omega ^{(s)} = \frac{{i\sqrt 2 }}{{{\omega _s}}}\frac{{g_\omega ^{(1)}g_\omega ^{(2)}}}{{\sqrt {g_\omega ^{{{(1)}^2}} + g_\omega ^{{{(2)}^2}}} }}{C_s}(t)
\end{equation}

\begin{equation}
\label{eq58}
\dot \mu _\omega ^{(a)} + 2i{\Delta _a}\mu _\omega ^{(a)} = \frac{{i\sqrt 2 }}{{{\omega _a}}}\frac{{g_\omega ^{(1)}g_\omega ^{(2)}}}{{\sqrt {g_\omega ^{{{(1)}^2}} + g_\omega ^{{{(2)}^2}}} }}{C_a}(t)
\end{equation}
where ${\Delta _s} = \omega - {\omega _s} $, ${\Delta _a} = \omega - {\omega _a}$ and we use that ${\omega ^2} - \omega _s^2 \approx  2{\Delta _s}{\omega _s}$ and ${\omega ^2} - \omega _a^2 \approx  2{\Delta _a}{\omega _a}$.

Further, we integrate Eqs.~\eqref{eq57} and~\eqref{eq58} and get
\begin{equation}
\label{eq59}
\mu _\omega ^{(s)}\left( t \right) = \mu _\omega ^{(s)}\left( 0 \right){e^{ - 2i{\Delta _s}t}} + \frac{{i\sqrt 2 }}{{{\omega _s}}}\frac{{g_\omega ^{(1)}g_\omega ^{(2)}}}{{\sqrt {g_\omega ^{{{(1)}^2}} + g_\omega ^{{{(2)}^2}}} }}\int\limits_0^t {{C_s}(\tau ){e^{ - 2i{\Delta _s}(t - \tau )}}d\tau }
\end{equation}
and a similar equation for $\mu _\omega ^{(a)}$.
To calculate the integral on the right side of Eq.~\eqref{eq59} we use Born approximation:
\begin{equation}
\label{eq60}
\int\limits_0^t {{C_s}(\tau ){e^{ - 2i{\Delta _s}(t - \tau )}}d\tau }  \approx {C_s}(t)\int\limits_0^\infty  {{e^{2i{\Delta _s}\tau }}d\tau }  = {C_s}(t)\int\limits_0^\infty  {{e^{ - 2i\left( {\omega_s  - {\omega}} \right)\,\tau '}}d\tau '}
\end{equation}
Using the Sokhotski-Plemelj formula \cite{ref74} we obtain that
\begin{equation}
\label{eq61}
\mu _\omega ^{(s)}\left( t \right) = \mu _\omega ^{(s)}\left( 0 \right){e^{ - 2i{\Delta _s}t}} + \frac{i}{{\sqrt 2 {\omega _s}}}\frac{{g_\omega ^{(1)}g_\omega ^{(2)}}}{{\sqrt {g_\omega ^{{{(1)}^2}} + g_\omega ^{{{(2)}^2}}} }}{C_s}(t)\left[ {\pi \,\delta \left( {\omega_s  - {\omega}} \right) + i\,{\rm{P}}\left( {\frac{1}{{\omega_s  - {\omega}}}} \right)} \right]
\end{equation}

\begin{equation}
\label{eq62}
\mu _\omega ^{(a)}\left( t \right) = \mu _\omega ^{(a)}\left( 0 \right){e^{ - 2i{\Delta _a}t}} + \frac{i}{{\sqrt 2 {\omega _a}}}\frac{{g_\omega ^{(1)}g_\omega ^{(2)}}}{{\sqrt {g_\omega ^{{{(1)}^2}} + g_\omega ^{{{(2)}^2}}} }}{C_a}(t)\left[ {\pi \,\delta \left( {\omega_a  - {\omega}} \right) + i\,{\rm{P}}\left( {\frac{1}{{\omega_a  - {\omega}}}} \right)} \right]
\end{equation}
We substitute these formulas into the equations for the slow amplitudes of the oscillators~\eqref{eq55} and~\eqref{eq56} and obtain:
\begin{equation}
\label{eq63}
\begin{array}{l}
{{\dot C}_s} = \frac{i}{{{\omega _s}}}\sum\limits_\omega  {g_\omega ^{\left( {ss} \right)}\left( {\mu _\omega ^{(s)}\left( 0 \right){e^{ - 2i{\Delta _s}t}} + \frac{i}{{\sqrt 2 {\omega _s}}}\frac{{g_\omega ^{(1)}g_\omega ^{(2)}}}{{\sqrt {g_\omega ^{{{(1)}^2}} + g_\omega ^{{{(2)}^2}}} }}{C_s}(t)\left[ {\pi \,\delta \left( {\omega_s  - {\omega}} \right) + i\,{\rm{P}}\left( {\frac{1}{{\omega_s  - {\omega}}}} \right)} \right]} \right)}  + \\
g_\omega ^{\left( {sa} \right)}\left( {\mu _\omega ^{(a)}\left( 0 \right){e^{ - 2i{\Delta _a}t}} + \frac{i}{{\sqrt 2 {\omega _a}}}\frac{{g_\omega ^{(1)}g_\omega ^{(2)}}}{{\sqrt {g_\omega ^{{{(1)}^2}} + g_\omega ^{{{(2)}^2}}} }}{C_a}(t)\left[ {\pi \,\delta \left( {\omega_a  - {\omega}} \right) + i\,{\rm{P}}\left( {\frac{1}{{\omega_a  - {\omega}}}} \right)} \right]} \right){e^{i\left( {{\omega _s} - {\omega _a}} \right)t}}
\end{array}
\end{equation}
and analogous equation for $\dot C_a (t)$.

We transform these equations in following way:
\begin{equation}
\label{eq65}
\begin{array}{l}
{{\dot C}_s} = f_1^{\left( 1 \right)}\left( t \right) + f_2^{\left( 1 \right)}\left( t \right) - \\
{C_s}\left( {\sum\limits_\omega  {\left\{ {\frac{{\pi \,(\rho _\omega ^{(1)}g_\omega ^{{{(1)}^2}} + \rho _\omega ^{(2)}g_\omega ^{{{(2)}^2}})}}{{4\omega _s^2}}\delta \left( {\omega_s  - {\omega}} \right)} \right\} + i\sum\limits_\omega  {\left\{ {\frac{{\rho _\omega ^{(1)}g_\omega ^{{{(1)}^2}} + \rho _\omega ^{(2)}g_\omega ^{{{(2)}^2}}}}{{4\omega _s^2}}\frac{1}{{\omega_s  - {\omega}}}} \right\}} } } \right) - \\
{C_a}\left( {\sum\limits_\omega  {\left\{ {\frac{{\pi \,(\rho _\omega ^{(1)}g_\omega ^{{{(1)}^2}} - \rho _\omega ^{(2)}g_\omega ^{{{(2)}^2}})}}{{4{\omega _s}{\omega _a}}}\delta \left( {\omega_a  - {\omega}} \right)} \right\} + i\sum\limits_\omega  {\left\{ {\frac{{\rho _\omega ^{(1)}g_\omega ^{{{(1)}^2}} - \rho _\omega ^{(2)}g_\omega ^{{{(2)}^2}}}}{{4{\omega _s}{\omega _a}}}\frac{1}{{\omega_a  - {\omega}}}} \right\}} } } \right){e^{i\left( {{\omega _s} - {\omega _a}} \right)t}}
\end{array}
\end{equation}

\begin{equation}
\label{eq66}
\begin{array}{l}
{{\dot C}_a} = f_1^{\left( 2 \right)}\left( t \right) + f_2^{\left( 2 \right)}\left( t \right) - \\
{C_s}\left( {\sum\limits_\omega  {\left\{ {\frac{{\pi \,(\rho _\omega ^{(1)}g_\omega ^{{{(1)}^2}} - \rho _\omega ^{(2)}g_\omega ^{{{(2)}^2}})}}{{4{\omega _s}{\omega _a}}}\delta \left( {\omega_s  - {\omega}} \right)} \right\} + i\sum\limits_\omega  {\left\{ {\frac{{\rho _\omega ^{(1)}g_\omega ^{{{(1)}^2}} + \rho _\omega ^{(2)}g_\omega ^{{{(2)}^2}}}}{{4{\omega _s}{\omega _a}}}\frac{1}{{\omega_s  - {\omega}}}} \right\}} } } \right){e^{i\left( {{\omega _a} - {\omega _s}} \right)t}} - \\
{C_a}\left( {\sum\limits_\omega  {\left\{ {\frac{{\pi \,(\rho _\omega ^{(1)}g_\omega ^{{{(1)}^2}} + \rho _\omega ^{(2)}g_\omega ^{{{(2)}^2}})}}{{4\omega _a^2}}\delta \left( {\omega_a  - {\omega}} \right)} \right\} + i\sum\limits_\omega  {\left\{ {\frac{{\rho _\omega ^{(1)}g_\omega ^{{{(1)}^2}} + \rho _\omega ^{(2)}g_\omega ^{{{(2)}^2}}}}{{4\omega _a^2}}\frac{1}{{\omega_a  - {\omega}}}} \right\}} } } \right)
\end{array}
\end{equation}
where $f_1^{\left( 1 \right)} = \frac{i}{{{\omega _s}}}\sum\limits_\omega  {g_\omega ^{\left( {ss} \right)}\mu _\omega ^{(s)}\left( 0 \right){e^{ - 2i{\Delta _s}t}}}$; $f_2^{\left( 1 \right)} = \frac{i}{{{\omega _s}}}\sum\limits_\omega  {g_\omega ^{\left( {sa} \right)}\mu _\omega ^{(a)}\left( 0 \right){e^{ - 2i{\Delta _s}t}}{e^{i({\omega _s} - {\omega _a})t}}}$; $f_1^{\left( 2 \right)} = \frac{i}{{{\omega _a}}}\sum\limits_\omega  {g_\omega ^{\left( {sa} \right)}\mu _\omega ^{(s)}\left( 0 \right){e^{ - 2i{\Delta _s}t}}{e^{-i({\omega _s} - {\omega _a})t}}}$; $f_2^{\left( 2 \right)} = \frac{i}{{{\omega _a}}}\sum\limits_\omega  {g_\omega ^{\left( {ss} \right)}\mu _\omega ^{(a)}\left( 0 \right){e^{ - 2i{\Delta _a}t}}}$ are noise terms.
We introduce notations: $\frac{{{\Gamma ^{\left(  +  \right)}}\left( {{\omega _{s,a}}} \right)}}{2} = \sum\limits_\omega  {\pi \,(\rho _\omega ^{(1)}g_\omega ^{{{(1)}^2}} + \rho _\omega ^{(2)}g_\omega ^{{{(2)}^2}})\delta \left( {\omega _{s,a}  - {\omega}} \right)}$;  $\frac{{{\Gamma ^{\left(  -  \right)}}\left( {{\omega _{s,a}}} \right)}}{2} = \\  \sum\limits_\omega  {\pi \,(\rho _\omega ^{(1)}g_\omega ^{{{(1)}^2}} - \rho _\omega ^{(2)}g_\omega ^{{{(2)}^2}})\delta \left( {\omega _{s,a}  - {\omega}} \right)}$; $\Delta \omega _{s,a}^{\left(  +  \right)} = \sum\limits_\omega  {\frac{{\rho _\omega ^{(1)}g_\omega ^{{{(1)}^2}} + \rho _\omega ^{(2)}g_\omega ^{{{(2)}^2}}}}{{\omega_{s,a}  - {\omega}}}}$; $\Delta \omega _{s,a}^{\left(  -  \right)} = \sum\limits_\omega  {\frac{{\rho _\omega ^{(1)}g_\omega ^{{{(1)}^2}} - \rho _\omega ^{(2)}g_\omega ^{{{(2)}^2}}}}{{\omega_{s,a}  - {\omega}}}}$. Here $\Gamma /2$ are the relaxation rates and  $\Delta \omega$ are the frequency shifts.

The expressions for $\Gamma /2$ and $\Delta \omega$ can be rewritten as
\begin{equation}
\label{eq67}
{\Gamma ^{\left(  \pm  \right)}}\left( {{\omega _{s,a}}} \right)/2 = V\pi \int {d\omega \,({\rho _1}(\omega )g_\omega ^{{{(1)}^2}} \pm {\rho _2}(\omega )g_\omega ^{{{(2)}^2}})\delta \left( {\omega_{s,a}  - {\omega}} \right)}
\end{equation}

\begin{equation}
\label{eq68}
\Delta \omega _{s,a}^{\left(  \pm  \right)} = {\rm P}.\int {d\omega \,\frac{{V({\rho _1}(\omega )g_\omega ^{{{(1)}^2}} \pm {\rho _2}(\omega )g_\omega ^{{{(2)}^2}})}}{{\omega_{s,a}  - {\omega}}}}
\end{equation}
where ${\rho _{1,2}}\left( \omega  \right)$ is a density of states in the respective reservoirs and $V$ is a reservoir volume.

Using the introduced notations, we rewrite Eqs.~\eqref{eq65} and~\eqref{eq66} as
\begin{equation}
\label{eq69}
\begin{array}{l}
{{\dot C}_s} = f_1^{\left( 1 \right)}\left( t \right) + f_2^{\left( 1 \right)}\left( t \right) - \\
{C_s}\left( {\left\{ {\frac{1}{{4\omega _s^2}}\frac{{{\Gamma ^{\left(  +  \right)}}\left( {{\omega _s}} \right)}}{2}} \right\} + i\left\{ {\frac{1}{{4\omega _s^2}}\Delta \omega _s^{\left(  +  \right)}} \right\}} \right) - \\
{C_a}{e^{i\left( {{\omega _s} - {\omega _a}} \right)t}}\left( {\left\{ {\frac{1}{{4{\omega _s}{\omega _a}}}\frac{{{\Gamma ^{\left(  -  \right)}}\left( {{\omega _a}} \right)}}{2}} \right\} + i\left\{ {\frac{1}{{4{\omega _s}{\omega _a}}}\Delta \omega _a^{\left(  -  \right)}} \right\}} \right)
\end{array}
\end{equation}

\begin{equation}
\label{eq70}
\begin{array}{l}
{{\dot C}_a} = f_1^{\left( 2 \right)}\left( t \right) + f_2^{\left( 2 \right)}\left( t \right) - \\
{C_s}{e^{i\left( {{\omega _a} - {\omega _s}} \right)t}}\left( {\left\{ {\frac{1}{{4{\omega _s}{\omega _a}}}\frac{{{\Gamma ^{\left(  -  \right)}}\left( {{\omega _s}} \right)}}{2}} \right\} + i\left\{ {\frac{1}{{4{\omega _s}{\omega _a}}}\Delta \omega _s^{\left(  -  \right)}} \right\}} \right) - \\
{C_a}\left( {\left\{ {\frac{1}{{4\omega _a^2}}\frac{{{\Gamma ^{\left(  +  \right)}}\left( {{\omega _a}} \right)}}{2}} \right\} + i\left\{ {\frac{1}{{4\omega _a^2}}\Delta \omega _a^{\left(  +  \right)}} \right\}} \right)
\end{array}
\end{equation}
Introducing notations
\begin{equation}
\label{eq71}
\beta _1^{\left( 1 \right)} = \left\{ {\frac{1}{{4\omega _s^2}}\frac{{{\Gamma ^{\left(  +  \right)}}\left( {{\omega _s}} \right)}}{2}} \right\} + i\left\{ {\frac{1}{{4\omega _s^2}}\Delta \omega _s^{\left(  +  \right)}} \right\}
\end{equation}

\begin{equation}
\label{eq72}
\beta _2^{\left( 1 \right)} = \left\{ {\frac{1}{{4{\omega _s}{\omega _a}}}\frac{{{\Gamma ^{\left(  -  \right)}}\left( {{\omega _a}} \right)}}{2}} \right\} + i\left\{ {\frac{1}{{4{\omega _s}{\omega _a}}}\Delta \omega _a^{\left(  -  \right)}} \right\}
\end{equation}

\begin{equation}
\label{eq73}
\beta _1^{\left( 2 \right)} = \left\{ {\frac{1}{{4{\omega _s}{\omega _a}}}\frac{{{\Gamma ^{\left(  -  \right)}}\left( {{\omega _s}} \right)}}{2}} \right\} + i\left\{ {\frac{1}{{4{\omega _s}{\omega _a}}}\Delta \omega _s^{\left(  -  \right)}} \right\}
\end{equation}

\begin{equation}
\label{eq74}
\beta _2^{\left( 2 \right)} = \left\{ {\frac{1}{{4\omega _a^2}}\frac{{{\Gamma ^{\left(  +  \right)}}\left( {{\omega _a}} \right)}}{2}} \right\} + i\left\{ {\frac{1}{{4\omega _a^2}}\Delta \omega _a^{\left(  +  \right)}} \right\}
\end{equation}
we write the Eqs.~\eqref{eq69} and~\eqref{eq70} as

\begin{equation}
\label{eq75}
{\dot C_s} = f_1^{\left( 1 \right)}\left( t \right) + f_2^{\left( 1 \right)}\left( t \right) - {C_s}\beta _1^{\left( 1 \right)} - {C_a}{e^{i\left( {{\omega _s} - {\omega _a}} \right)t}}\beta _2^{\left( 1 \right)}
\end{equation}

\begin{equation}
\label{eq76}
{\dot C_a} = f_1^{\left( 2 \right)}\left( t \right) + f_2^{\left( 2 \right)}\left( t \right) - {C_s}{e^{ - i\left( {{\omega _s} - {\omega _a}} \right)t}}\beta _1^{\left( 2 \right)} - {C_a}\,\beta _2^{\left( 2 \right)}
\end{equation}
Note that the real parts of $\beta$ are responsible for the relaxation processes and the imaginary parts of $\beta$ are responsible for the frequency shift.

Neglect the noise terms and accomplish the change of variable ${C_s} = {{\tilde C}_s}{e^{i\left( {{\omega _s} - {\omega _a}} \right)t/2}}$; ${C_a} = {{\tilde C}_a}{e^{ - i\left( {{\omega _s} - {\omega _a}} \right)t/2}}$ we obtain the following differential equation

\begin{equation}
\label{eq77}
\frac{d}{{dt}}\left( {\begin{array}{*{20}{c}}
{{{\tilde C}_s}}\\
{{{\tilde C}_a}}
\end{array}} \right) = \left( {\begin{array}{*{20}{c}}
{ - \beta _1^{(1)} - i\frac{{{\omega _s} - {\omega _a}}}{2}}&{ - \beta _2^{(1)}}\\
{ - \beta _1^{(2)}}&{ - \beta _2^{(2)} + i\frac{{{\omega _s} - {\omega _a}}}{2}}
\end{array}} \right)\left( {\begin{array}{*{20}{c}}
{{{\tilde C}_s}}\\
{{{\tilde C}_a}}
\end{array}} \right)
\end{equation}
We introduce new variables: ${{\tilde C}_s} = \left( {{a_1} + {a_2}} \right)/\sqrt 2$ and ${{\tilde C}_a} = \left( {{a_1} - {a_2}} \right)/\sqrt 2$ (then ${x_{1,2}} = {a_{1,2}}{e^{i{\omega _0}t}} + c.c$) and rewrite Eq.~\eqref{eq77} as

\begin{equation}
\label{eq78}
\frac{d}{{dt}}\left( {\begin{array}{*{20}{c}}
{{a_1}}\\
{{a_2}}
\end{array}} \right) =  - \left( {\begin{array}{*{20}{c}}
{\frac{{\beta _1^{\left( 1 \right)} + \beta _2^{\left( 2 \right)}}}{2} + \frac{{\beta _2^{\left( 1 \right)} + \beta _1^{\left( 2 \right)}}}{2}}&{\frac{{\beta _1^{\left( 1 \right)} - \beta _2^{\left( 2 \right)}}}{2} - \frac{{\beta _2^{\left( 1 \right)} - \beta _1^{\left( 2 \right)}}}{2} + i\frac{{{\kappa ^2}}}{{2{\omega _0}}}}\\
{\frac{{\beta _1^{\left( 1 \right)} - \beta _2^{\left( 2 \right)}}}{2} + \frac{{\beta _2^{\left( 1 \right)} - \beta _1^{\left( 2 \right)}}}{2} + i\frac{{{\kappa ^2}}}{{2{\omega _0}}}}&{\frac{{\beta _1^{\left( 1 \right)} + \beta _2^{\left( 2 \right)}}}{2} - \frac{{\beta _2^{\left( 1 \right)} + \beta _1^{\left( 2 \right)}}}{2}}
\end{array}} \right)\left( {\begin{array}{*{20}{c}}
{{a_1}}\\
{{a_2}}
\end{array}} \right)
\end{equation}
where we use that ${\omega _s} - {\omega _a} \approx {\kappa ^2}/{\omega _0}$.

Using the expressions~\eqref{eq71}-\eqref{eq74} we obtain the following system of equations

\begin{equation}
\label{eq79}
\begin{array}{l}
\frac{d}{{dt}}\left( {\begin{array}{*{20}{c}}
{{a_1}}\\
{{a_2}}
\end{array}} \right) = \\
 - \frac{{\pi V}}{{4\omega _0^2}}\left( {\begin{array}{*{20}{c}}
{{\rho _1}({\omega _s})g_{{\omega _s}}^{{{(1)}^2}} + {\rho _1}({\omega _a})g_{{\omega _a}}^{{{(1)}^2}}}&{{\rho _1}({\omega _s})g_{{\omega _s}}^{{{(1)}^2}} - {\rho _1}({\omega _a})g_{{\omega _a}}^{{{(1)}^2}} + \frac{{4\omega _0^2}}{{\pi V}}\frac{{i{\kappa ^2}}}{{2{\omega _0}}}}\\
{{\rho _2}({\omega _s})g_{{\omega _s}}^{{{(2)}^2}} - {\rho _2}({\omega _a})g_{{\omega _a}}^{{{(2)}^2}} + \frac{{4\omega _0^2}}{{\pi V}}\frac{{i{\kappa ^2}}}{{2{\omega _0}}}}&{{\rho _2}({\omega _s})g_{{\omega _s}}^{{{(2)}^2}} + {\rho _2}({\omega _a})g_{{\omega _a}}^{{{(2)}^2}}}
\end{array}} \right)\left( {\begin{array}{*{20}{c}}
{{a_1}}\\
{{a_2}}
\end{array}} \right)
\end{array}
\end{equation}
Introducing new quantities $\Omega  = {\kappa ^2}/2{\omega _0}$ and $\gamma _\omega ^{(1),(2)} = V\,g_{{\omega _{s,a}}}^{(1),(2)}/2{\omega _0}$, we can rewrite Eq.~\eqref{eq79} as

\begin{equation}
\label{eq80}
\frac{d}{{dt}}\left( {\begin{array}{*{20}{c}}
{{a_1}}\\
{{a_2}}
\end{array}} \right) =  - \left( {\begin{array}{*{20}{c}}
{\frac{{{\gamma _1}({\omega _s}) + {\gamma _1}({\omega _a})}}{2}}&{\frac{{{\gamma _1}({\omega _s}) - {\gamma _1}({\omega _a})}}{2} + i\Omega }\\
{\frac{{{\gamma _2}({\omega _s}) - {\gamma _2}({\omega _a})}}{2} + i\Omega }&{\frac{{{\gamma _2}({\omega _s}) + {\gamma _2}({\omega _a})}}{2}}
\end{array}} \right)\left( {\begin{array}{*{20}{c}}
{{a_1}}\\
{{a_2}}
\end{array}} \right)
\end{equation}
where ${\gamma _{1,2}}\left( {{\omega _{s,a}}} \right) = \pi \,{\rho _{1,2}}({\omega _{s,a}}){\left( {\gamma _{{\omega _{s,a}}}^{(1),(2)}} \right)^2}$ are the relaxation rates of the first and second oscillators at the frequencies of symmetric and anti-symmetric eigenmodes, respectively.

In Eq.~\eqref{eq80}, we can approximately assume that $\left( {{\gamma _{1,2}}({\omega _s}) + {\gamma _{1,2}}({\omega _a})} \right)/2 \approx {\gamma _{1,2}}({\omega _0})$ and introduce designation ${\rm K}\left( {\kappa ,\,{\gamma _{1,2}}} \right) = \frac{{{\gamma _{1,2}}({\omega _s}) - {\gamma _{1,2}}({\omega _a})}}{2}$. It is seen that Eq.~\eqref{eq80} are equivalent to Eq.~\eqref{eq47} after transition to slowly varying amplitudes.

\section{Dynamics of the density matrix of two coupled oscillators}

We simulate the quantum dynamics of two coupled oscillators predicted by the master equation for the density matrix~\eqref{eq41}. We consider that at the initial time the first oscillator is in the state with one excitation and the second oscillator is in the ground state, i.e. the system state is $\left| {\psi \left( {t = 0} \right)} \right\rangle  = \left| {1,\,0} \right\rangle$. Also we assume that the temperatures of two reservoirs are zero. In these conditions, the system evolution takes place in the subspace spanned by the states: $\left| {1,\,0} \right\rangle $, $\left| {0,\,1} \right\rangle $ and $\left| {0,\,0} \right\rangle $.

\begin{figure}[t]
  \centering
 \includegraphics[width=\linewidth]{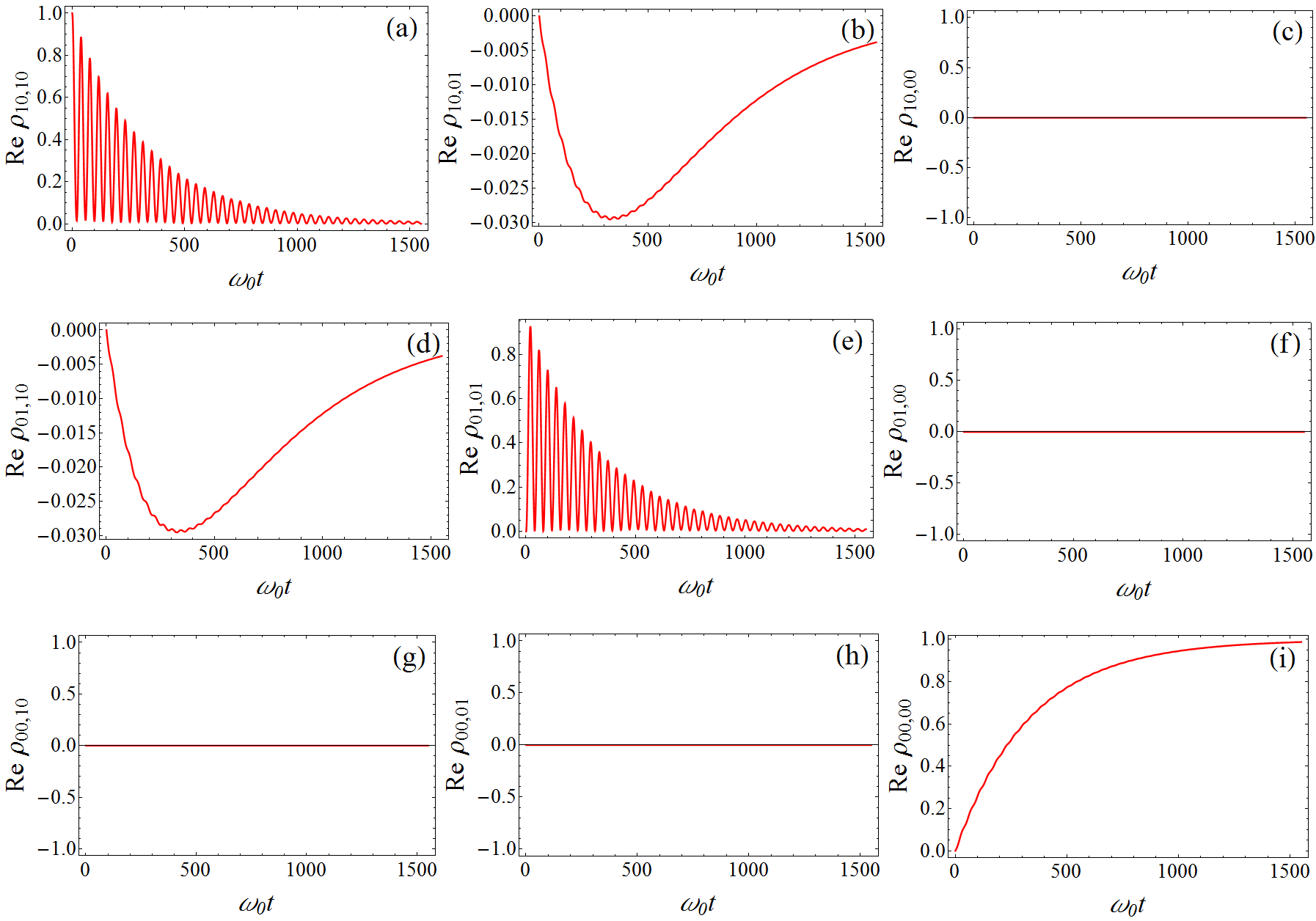}
  \caption{Dependence of the real parts of the elements of the density matrix in the EASC regime. $\rho \left( \omega  \right) \sim {\omega ^2}$; $\gamma_1 = 0.001 \omega_0$; $\gamma_2 = 0.002 \omega_0$; $\Omega = 0.08 \omega_0$.}
  \label{fig:figure1A}
\end{figure}

\begin{figure}[t]
  \centering
 \includegraphics[width=\linewidth]{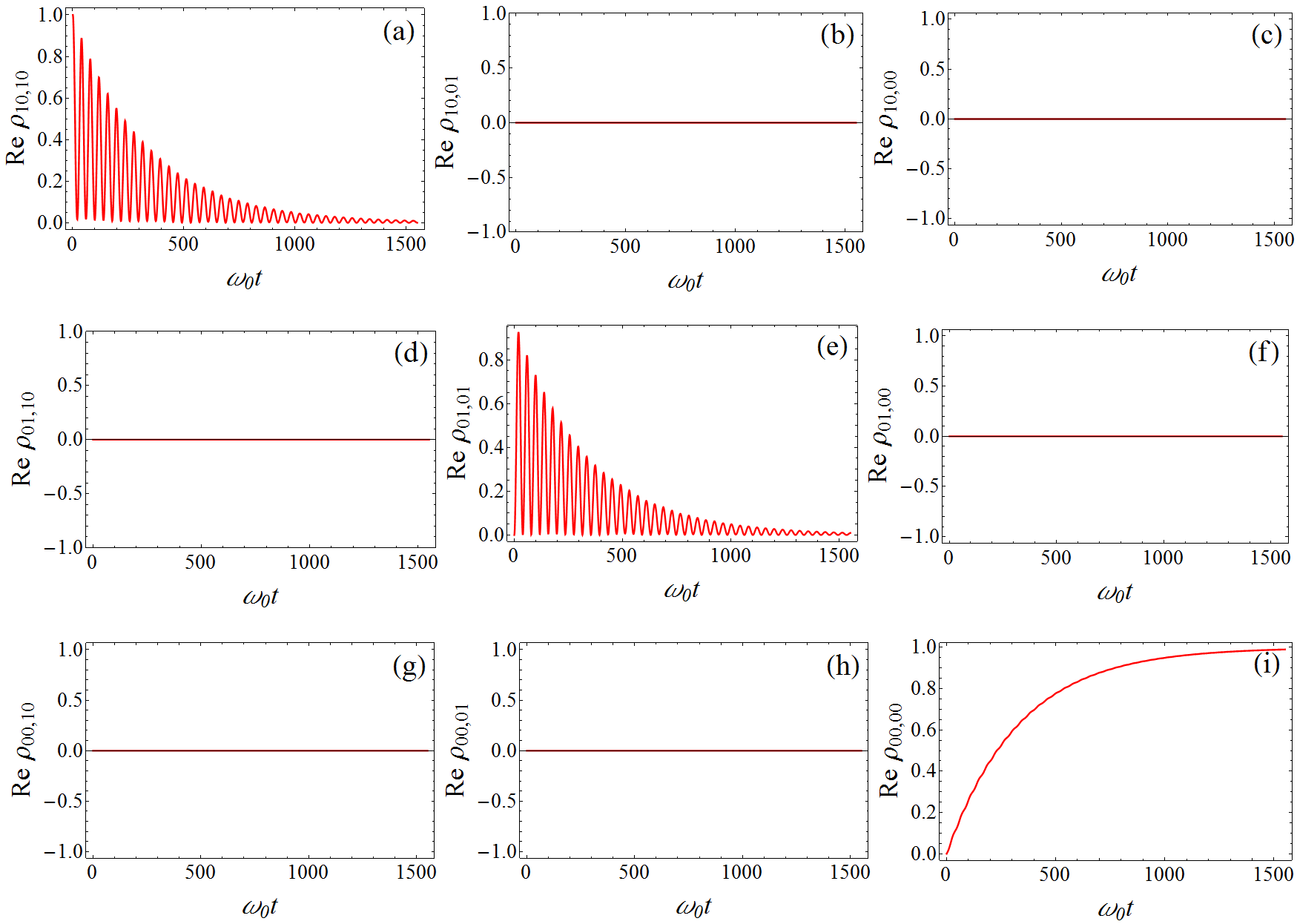}
  \caption{Dependence of the real parts of the elements of the density matrix in the SC regime. $\rho \left( \omega  \right) = const$; $\gamma_1 = 0.001 \omega_0$; $\gamma_2 = 0.002 \omega_0$; $\Omega = 0.08 \omega_0$.}
  \label{fig:figure2A}
\end{figure}

We calculate the temporal evolution of the density matrix of the system with the power-dependent reservoir density of states $\rho \left( \omega  \right) \sim {\omega ^2}$ (Fig.~\ref{fig:figure1A}) and with the frequency-independent reservoirs $\rho \left( \omega  \right) = const$ (Fig.~\ref{fig:figure2A}). In the first system, the EI coupling strength is nonzero. While in the second system the EI coupling is zero.

The EI coupling leads to change in the system dynamics (cf. Figs.~\ref{fig:figure1A} and~\ref{fig:figure2A}). The influence of EI coupling is most pronounced in the time dependence of the real parts of the non-diagonal elements of density matrix ${\hat \rho _{10,01}}$ and ${\hat \rho _{01,10}}$. Our calculations show that in the case of frequency-independent reservoirs, these real parts are zero (Fig.~\ref{fig:figure2A}b, d). While in the case of the power-dependent reservoir density of states its differ from zero during the system evolution (Fig.~\ref{fig:figure1A}b, d). This leads to the nonzero interaction energy $\Omega \left\langle {\hat a_1^\dag {{\hat a}_2} + {{\hat a}_1}\hat a_2^\dag } \right\rangle $ during the system evolution (Fig.~\ref{fig:figure5} in the main text).

\begin{figure}[t]
  \centering
 \includegraphics[width=0.7\linewidth]{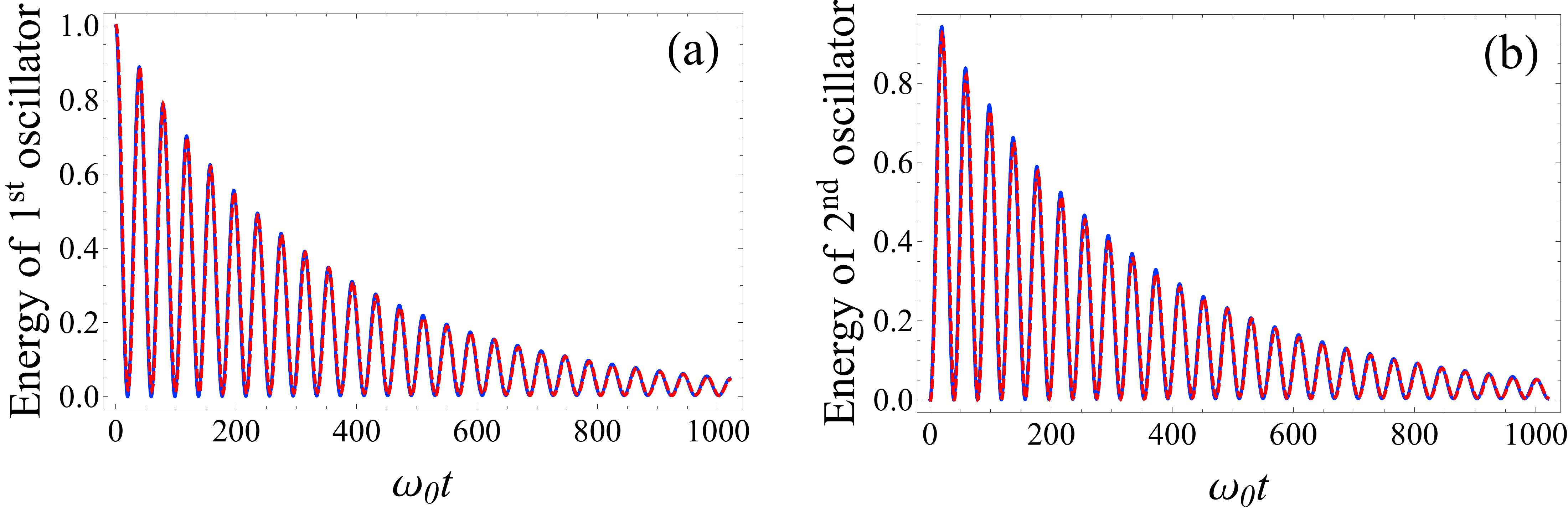}
  \caption{Dependence of the energy of the first (a) and the second (b) oscillators calculated from Eq.~\eqref{eq3}, ${E_{1,2}} = {\left| {{a_{1,2}}} \right|^2}$, and from the master equation for density matrix~\eqref{eq41}, ${E_{1,2}} = \left\langle {\hat a_{1,2}^\dag {{\hat a}_{1,2}}} \right\rangle$, in the EASC regime. $\rho \left( \omega  \right) \sim {\omega ^2}$; $\gamma_1 = 0.001 \omega_0$; $\gamma_2 = 0.002 \omega_0$; the Rabi coupling strength $\Omega = 0.08 \omega_0$.}
  \label{fig:figure4A}
\end{figure}

In addition, we compare the time dependencies of the oscillator's energies calculated from Eq.~\eqref{eq3}, ${E_{1,2}} = {\left| {{a_{1,2}}} \right|^2}$, and from the master equation for density matrix~\eqref{eq41}, ${E_{1,2}} = \left\langle {\hat a_{1,2}^\dag {{\hat a}_{1,2}}} \right\rangle$. It is seen that Eq.~\eqref{eq3} and the master equations for the density matrix~\eqref{eq41} lead to identical time dependencies in all regimes, including EASC regime (Fig.~\ref{fig:figure4A}).

Thus, we conclude that the EI coupling has a noticeable effect in the quantum limit.

\section{Influence of the counter-rotating and diamagnetic terms on the environmental - induced coupling}

Usually, the transition to the ultra-strong coupling regime is determined by the manifestation of influence of the counter-rotating and diamagnetic terms on the system behavior \cite{ref61}. The Hamiltonian of system of two coupled oscillators taking into account these terms can be presented in the form \cite{ref61}

\begin{equation}
\label{eq81}
{\hat H_S} = {\hat H_{JC}} + {\hat H_{CRW}} + {\hat H_{dia}}
\end{equation}
Here ${\hat H_{JC}} = {\omega _0}\,\hat a_1^\dag {\hat a_1} + {\omega _0}\,\hat a_2^\dag {\hat a_2} + \Omega \left( {\hat a_1^\dag {{\hat a}_2} + {{\hat a}_1}\hat a_2^\dag } \right)$ is the Jaynes-Cumming Hamiltonian [10], ${\hat H_{CRW}} = \Omega \left( {\hat a_1^\dag \hat a_2^\dag  + {{\hat a}_1}{{\hat a}_2}} \right)$ is the counter-rotating term, ${\hat H_{dia}} = {D_1}{\left( {\hat a_1^\dag  + {{\hat a}_1}} \right)^2} + {D_2}{\left( {\hat a_2^\dag  + {{\hat a}_2}} \right)^2}$ is the diamagnetic terms where ${D_{1,2}} \ge {\Omega ^2}/{\omega _0}$ \cite{ref61}. For simplicity, we consider the case ${D_1} = {D_2} = {\Omega ^2}/{\omega _0}$.

Using the Heisenberg equations, we obtain the equations for average values of operators $\left\langle {{a_{1,2}}} \right\rangle$ and $\left\langle {\hat a_{1,2}^ + } \right\rangle$ 

\begin{equation}
\label{eq82}
\begin{array}{l}
\frac{d}{{dt}}\left( {\begin{array}{*{20}{c}}
{\left\langle {{a_1}} \right\rangle }\\
{\left\langle {{a_2}} \right\rangle }\\
{\left\langle {a_1^ + } \right\rangle }\\
{\left\langle {a_2^ + } \right\rangle }
\end{array}} \right) = \\
\left( {\begin{array}{*{20}{c}}
{ - i\,{\omega _0} - 2i\,{\Omega ^2}/{\omega _0}}&{ - i\,\Omega }&{ - 2i\,{\Omega ^2}/{\omega _0}}&{ - i\,\Omega }\\
{ - i\,\Omega }&{ - i\,\omega_0  - 2i\,{\Omega ^2}/{\omega _0}}&{ - i\,\Omega }&{ - 2i\,{\Omega ^2}/{\omega _0}}\\
{2i\,{\Omega ^2}/{\omega _0}}&{i\,\Omega }&{i\,\omega_0  + 2i\,{\Omega ^2}/{\omega _0}}&{i\,\Omega }\\
{i\,\Omega }&{2i\,{\Omega ^2}/{\omega _0}}&{i\,\Omega }&{i\,\omega_0  + 2i\,{\Omega ^2}/{\omega _0}}
\end{array}} \right)\left( {\begin{array}{*{20}{c}}
{\left\langle {{a_1}} \right\rangle }\\
{\left\langle {{a_2}} \right\rangle }\\
{\left\langle {a_1^ + } \right\rangle }\\
{\left\langle {a_2^ + } \right\rangle }
\end{array}} \right)
\end{array}
\end{equation}
Without taking into account the counter-rotating and diamagnetic terms the equations for $\left\langle {{a_{1,2}}} \right\rangle $  and $\left\langle {\hat a_{1,2}^ + } \right\rangle $ are divided into two independent subsystems:

\begin{equation}
\label{eq83}
\frac{d}{{dt}}\left( {\begin{array}{*{20}{c}}
{\left\langle {{a_1}} \right\rangle }\\
{\left\langle {{a_2}} \right\rangle }
\end{array}} \right) = \left( {\begin{array}{*{20}{c}}
{ - i\,{\omega _0}}&{ - i\,\Omega }\\
{ - i\,\Omega }&{ - i\,{\omega _0}}
\end{array}} \right)\left( {\begin{array}{*{20}{c}}
{\left\langle {{a_1}} \right\rangle }\\
{\left\langle {{a_2}} \right\rangle }
\end{array}} \right)
\end{equation}
and similar for $\left\langle {\hat a_{1,2}^ + } \right\rangle $. The interaction of the system with the environment leads to the relaxation of the oscillator amplitudes, i.e. $\left\langle {{a_{1,2}}} \right\rangle $. The relaxation processes depend on the eigenstates of coupled system (see Appendices A and B). The eigenstates of coupled system described by Eq.~\eqref{eq83} are symmetric and anti-symmetric modes. The difference of the eigenstates from the states of individual oscillators leads to the appearance of the environmental-induced coupling (see Appendices A and B).

\begin{figure}[t]
  \centering
 \includegraphics[width=1\linewidth]{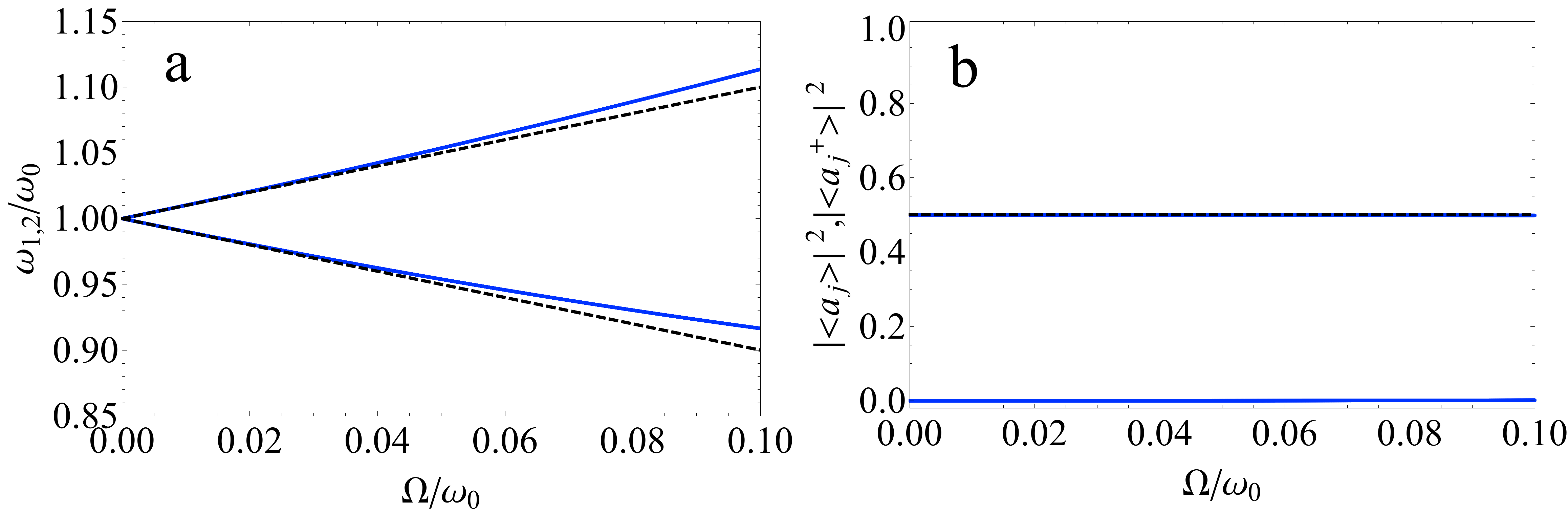}
  \caption{(a) Dependence of eigenfrequencies of Eq.~\eqref{eq82} (solid blue lines) and Eq.~\eqref{eq83} (dashed black lines) on the coupling strength. (b) The dependence of the components of eigenstates of Eq.~\eqref{eq82} ($\left\langle {{a_1}} \right\rangle$, $\left\langle {{a_2}} \right\rangle$, $\left\langle {a_1^ + } \right\rangle$, $\left\langle {a_2^ + } \right\rangle$) (solid blue lines) and Eq.~\eqref{eq83} ($\left\langle {{a_1}} \right\rangle$, $\left\langle {{a_2}} \right\rangle$) (dashed black lines) on the coupling strength $\Omega$.}
  \label{fig:figure3A}
\end{figure}

The eigenstates and eigenfrequencies of the system of equations~\eqref{eq82} taking into account the counter-rotating and diamagnetic terms differ from ones of the system of equations~\eqref{eq83}. This leads, among other things, to a change in the relaxation processes in the system. However, the changes in the eigenstates and eigenfrequencies caused by the counter-rotating and diamagnetic terms become significant only when the coupling strength $\Omega$ is at least greater than $0.1\,{\omega _0}$ \cite{ref61,ref70}. Indeed, the eigenstates of Eq.~\eqref{eq83} are symmetric and anti-symmetric states for all values of the coupling strength. The counter-rotating and diamagnetic terms modify the eigenstates but the corresponding change is proportional to $\Omega /{\omega _0}$  and is not significant when $\Omega /{\omega _0} <  < 1$, see Fig.~\ref{fig:figure3A}. The eigenfrequencies of symmetric and anti-symmetric eigenstates of the system~\eqref{eq83} are ${\omega _{S,A}} = {\omega _0} \pm \Omega$, while the corresponding eigenfrequencies of the system~\eqref{eq82} are ${\tilde \omega _{S,A}} = \sqrt {\omega _0^2 \pm 2\Omega {\omega _0} + 4{\Omega ^2}}  \approx {\omega _0} \pm \Omega  + \frac{{3{\Omega ^2}}}{{2{\omega _0}}} + ...$. It is seen that the change in the eigenfrequencies is proportional to ${\Omega ^2}/{\omega _0}$ and becomes comparable with the frequency splitting ($2\,\Omega$) only when $\Omega /{\omega _0} \sim 1$, see Fig.~\ref{fig:figure3A}.

Thus, the changes in the eigenfrequencies and eigenstates caused by the counter-rotating and diamagnetic terms are of the order of $\Omega /{\omega _0}$ compared to the changes related to the Jaynes-Cumming interaction, i.e. $\Omega \left( {\hat a_1^\dag {{\hat a}_2} + {{\hat a}_1}\hat a_2^\dag } \right)$. For this reason, the changes in the relaxation processes caused by the counter-rotating and diamagnetic terms are also of the order of $\Omega /{\omega _0}$  compared to changes in the relaxation processes related to the Jaynes-Cumming interaction. 

\bibliographystyle{plain}

\begin{thebibliography}{9}

\bibitem{ref1}
  R. Chikkaraddy, B. De Nijs, F. Benz, et al. Single-molecule strong coupling at room temperature in plasmonic nanocavities,
  \href{https://doi.org/10.1038/nature17974}{Nature
        \textbf{535}, 127-130 (2016).}

\bibitem{ref2}
  T. Hummer, F. Garcia-Vidal, L. Martin-Moreno, D. Zueco. Weak and strong coupling regimes in plasmonic QED,
  \href{https://doi.org/10.1103/PhysRevB.87.115419}{Phys. Rev. B \textbf{87}, 115419 (2013).}

\bibitem{ref3}
  P. Torma, W. L. Barnes. Strong coupling between surface plasmon polaritons and emitters: a review,
  \href{https://doi.org/10.1088/0034-4885/78/1/013901}{Rep. Prog. Phys. \textbf{78}, 013901 (2015).}

\bibitem{ref4}
  G. Zengin, M. Wersall, S. Nilsson, T. J. Antosiewicz, M. Kall, T. Shegai. Realizing Strong Light-Matter Interactions between Single-Nanoparticle Plasmons and Molecular Excitons at Ambient Conditions,
  \href{https://doi.org/10.1103/PhysRevLett.114.157401}{Phys. Rev. Lett. \textbf{114}, 157401 (2015).}

\bibitem{ref5}
  B. Munkhbat, M. Wersall, D. G. Baranov, T. J. Antosiewicz, T. Shegai. Suppression of photo-oxidation of organic chromophores by strong coupling to plasmonic nanoantennas,
  \href{https://doi.org/10.1126/sciadv.aas9552}{Sci. Adv. \textbf{4}, eaas9552 (2018).}

\bibitem{ref6}
  T. Yoshie, A. Scherer, J. Hendrickson, G. Khitrova, H. Gibbs, G. Rupper, C. Ell, O. Shchekin, D. Deppe. Vacuum Rabi splitting with a single quantum dot in a photonic crystal nanocavity,
  \href{https://doi.org/10.1038/nature03119}{Nature \textbf{432}, 200 (2004).}

\bibitem{ref7}
  F. Benz, M.K. Schmidt, A. Dreismann, et al. Single-molecule optomechanics in “picocavities,
  \href{https://doi.org/10.1126/science.aah5243}{Science \textbf{354}(6313), 726-729 (2016).}

\bibitem{ref8}
 K. Hennessy, A. Badolato, M. Winger, D. Gerace, M. Atatüre, S. Gulde, S. Fält, E. L. Hu, A. Imamoğlu. Quantum nature of a strongly coupled single quantum dot–cavity system,
  \href{https://doi.org/10.1038/nature05586}{Nature \textbf{445}, 896 (2007).}

\bibitem{ref9}
  J. P. Reithmaier,  G. Sęk, A. Löffler, et al. Strong coupling in a single quantum dot–semiconductor microcavity system,
  \href{https://doi.org/10.1038/nature02969}{Nature \textbf{432}(7014), 197-200 (2004).}

\bibitem{ref10}
  S. Zanotto, A. Tredicucci. Universal lineshapes at the crossover between weak and strong critical coupling in Fano-resonant coupled oscillators,
  \href{https://doi.org/10.1038/srep24592}{Sci. Rep. \textbf{6}, 24592 (2016).}

\bibitem{ref11}
  S. Zanotto, F. P. Mezzapesa, F. Bianco, G. Biasiol, L. Baldacci, M. S. Vitiello, L. Sorba, R. Colombelli, A. Tredicucci. Perfect energy-feeding into strongly coupled systems and interferometric control of polariton absorption,
  \href{https://doi.org/10.1038/nphys3106}{Nature Phys. \textbf{10}, 830 (2014).}

\bibitem{ref12}
  M.A. Miri, A. Alu. Exceptional points in optics and photonics,
  \href{https://doi.org/10.1126/science.aar7709}{Science \textbf{363}, 6422 (2019).}

\bibitem{ref13}
  S. K. Ozdemir, S. Rotter, F. Nori, L. Yang. Parity–time symmetry and exceptional points in photonics,
  \href{https://doi.org/10.1038/s41563-019-0304-9}{Nature Mater. \textbf{18}, 783 (2019).}

\bibitem{ref14}
  W. Gao, X. Li, M. Bamba, J. Kono. Continuous transition between weak and ultrastrong coupling through exceptional points in carbon nanotube microcavity exciton–polaritons,
  \href{https://doi.org/10.1038/s41566-018-0157-9}{Nature Photon. \textbf{12}, 362 (2018).}

\bibitem{ref82}
B. Peng, Ş. K. Özdemir, S. Rotter, H. Yilmaz, M. Liertzer, F. Monifi, C. M. Bender, F. Nori, L. Yang. Loss-induced suppression and revival of lasing, \href{https://doi.org/10.1126/science.1258004}{Science \textbf{346}, 328-332 (2014).}

\bibitem{ref15}
  N. Moiseyev. 
  \href{https://ui.adsabs.harvard.edu/abs/2011nhqm.book.....M/abstract}{Non-Hermitian quantum mechanics, Cambridge University Press (2011).}

\bibitem{ref16}
  M.V. Berry. Physics of Nonhermitian Degeneracies,
  \href{https://doi.org/10.1023/B:CJOP.0000044002.05657.04}{Czech. J. Phys. \textbf{54}, 1039 (2004).}

\bibitem{ref17}
  D. Sanvitto, S. Kena-Cohen. The road towards polaritonic devices,
  \href{https://doi.org/10.1038/nmat4668}{Nature Mater. \textbf{15}, 1061 (2016).}

\bibitem{ref18}
  D.D. Smith, H. Chang, K.A. Fuller, A. Rosenberger, R.W. Boyd. Coupled-resonator-induced transparency,
  \href{https://doi.org/10.1103/PhysRevA.69.063804}{Phys. Rev. A \textbf{69}, 063804 (2004).}
  
  \bibitem{ref19}
  E. Orgiu, J. George, J.A. Hutchison, et al. Conductivity in organic semiconductors hybridized with the vacuum field,
  \href{https://doi.org/10.1038/nmat4392}{Nature Mater. \textbf{14}, 1123-1129 (2015).}

\bibitem{ref20}
  J.A. Hutchison, A. Liscio, T. Schwartz, A. Canaguier-Durand, C. Genet, V. Palermo, P. Samori, T.W. Ebbesen. Tuning the Work-Function Via Strong Coupling,
  \href{https://doi.org/10.1002/adma.201203682}{Adv. Mater. \textbf{25}, 2481 (2013).}

\bibitem{ref21}
  J.A. Hutchison, T. Schwartz, C. Genet, E. Devaux, T.W. Ebbesen. Modifying Chemical Landscapes by Coupling to Vacuum Fields,
  \href{https://doi.org/10.1002/anie.201107033}{Angew. Chem. Int. Ed. \textbf{51}, 1592 (2012).}
 
\bibitem{ref22}
  J. Galego, F.J. Garcia-Vidal, J. Feist. Suppressing photochemical reactions with quantized light fields,
  \href{https://doi.org/10.1038/ncomms13841}{Nat. Commun. 
        \textbf{7}, 13841 (2016).}

\bibitem{ref23}
  N. Nefedkin, E. Andrianov, A. Vinogradov. The role of strong coupling in the process of photobleaching suppression,
  \href{https://doi.org/10.1021/acs.jpcc.0c05518}{J. Phys. Chem. C \textbf{124}, 18234 (2020).}

\bibitem{ref24}
 A.A. Zyablovsky, I.V. Doronin, E.S. Andrianov, A.A. Pukhov, Y.E. Lozovik, A. P. Vinogradov, A.A. Lisyansky. Exceptional Points as Lasing Prethresholds,
  \href{https://doi.org/10.1002/lpor.202000450}{Laser Photonics Rev. \textbf{15}, 2000450 (2021).}

\bibitem{ref25}
  I.V. Doronin, A.A. Zyablovsky, E.S. Andrianov. Strong-coupling-assisted formation of coherent radiation below the lasing threshold,
  \href{https://doi.org/10.1364/OE.417354}{Opt. Express \textbf{29}, 5624 (2021).}

\bibitem{ref26}
  I.V. Doronin, A.A. Zyablovsky, E.S. Andrianov, A.A. Pukhov, A.P. Vinogradov. Lasing without inversion due to parametric instability of the laser near the exceptional point,
  \href{https://doi.org/10.1103/PhysRevA.100.021801}{Phys. Rev. A \textbf{100}, 021801(R) (2019).}

\bibitem{ref27}
  X. Liu, T. Galfsky, Z. Sun, F. Xia, E.-C. Lin, Y.-H. Lee, S. Kena-Cohen, V.M. Menon. Strong light–matter coupling in two-dimensional atomic crystals,
  \href{https://doi.org/10.1038/nphoton.2014.304}{Nature Photon. \textbf{9}, 30 (2015).}

\bibitem{ref28}
  I.-C. Hoi, C. Wilson, G. Johansson, T. Palomaki, B. Peropadre, P. Delsing. Demonstration of a single-photon router in the microwave regime,
  \href{https://doi.org/10.1103/PhysRevLett.107.073601}{Phys. Rev. Lett. \textbf{107}, 073601 (2011).}

\bibitem{ref29}
 T. Volz, A. Reinhard, M. Winger, A. Badolato, K.J. Hennessy, E.L. Hu, A. Imamoğlu. Ultrafast all-optical switching by single photons,
  \href{https://doi.org/10.1038/nphoton.2012.181}{Nature Photon. \textbf{6}, 605 (2012).}

\bibitem{ref30}
  R. Schoelkopf, S. Girvin. Wiring up quantum systems,
  \href{https://doi.org/10.1038/451664a}{Nature \textbf{451}, 664 (2008).}

\bibitem{ref31}
  T.D. Ladd, F. Jelezko, R. Laflamme, Y. Nakamura, C. Monroe, J.L. O’Brien. Quantum computers,
  \href{https://doi.org/10.1038/nature08812}{Nature \textbf{464}, 45 (2010).}

\bibitem{ref32}
  Z.-L. Xiang, S. Ashhab, J. You, F. Nori. Hybrid quantum circuits: Superconducting circuits interacting with other quantum systems,
  \href{https://doi.org/10.1103/RevModPhys.85.623}{Rev. Mod. Phys. \textbf{85}, 623 (2013).}

\bibitem{ref33-}
  R. Hanai, A. Edelman, Y. Ohashi, P.B. Littlewood. Non-Hermitian phase transition from a polariton Bose-Einstein condensate to a photon laser,
  \href{https://doi.org/10.1103/PhysRevLett.122.185301}{Phys. Rev. Lett. \textbf{122}, 185301 (2019).}

\bibitem{ref34}
  Y.-H. Lai, Y.-K. Lu, M.-G. Suh, Z. Yuan, K. Vahala. Observation of the exceptional-point-enhanced Sagnac effect,
  \href{https://doi.org/10.1038/s41586-019-1777-z}{Nature \textbf{576}, 65 (2019).}

\bibitem{ref35}
  H. Hodaei, A.U. Hassan, S. Wittek, H. Garcia-Gracia, R. El-Ganainy, D.N. Christodoulides, M. Khajavikhan. Enhanced sensitivity at higher-order exceptional points,
  \href{https://doi.org/10.1038/nature23280}{Nature \textbf{548}, 187 (2017).}

\bibitem{ref36}
  W. Chen, S. K. Ozdemir, G. Zhao, J. Wiersig, L. Yang. Exceptional points enhance sensing in an optical microcavity,
  \href{https://doi.org/10.1038/nature23281}{Nature \textbf{548}, 192 (2017).}

\bibitem{ref37}
  J. Flick, N. Rivera, P. Narang. Strong light-matter coupling in quantum chemistry and quantum photonics,
  \href{https://doi.org/10.1515/nanoph-2018-0067}{Nanophotonics \textbf{7}, 1479 (2018).}

\bibitem{ref38}
  A. Thomas, L. Lethuillier-Karl, K. Nagarajan. Tilting a ground-state reactivity landscape by vibrational strong coupling,
  \href{https://doi.org/10.1126/science.aau7742}{Science \textbf{363}, 615 (2019).}

\bibitem{ref39}
  A. Faraon, I. Fushman, D. Englund, N. Stoltz, P. Petroff, J. Vučković. Coherent generation of non-classical light on a chip via photon-induced tunnelling and blockade,
  \href{https://doi.org/10.1038/nphys1078}{Nature Phys. \textbf{4}, 859 (2008).}
  
  \bibitem{ref40}
  H. Hodaei, M.-A. Miri, M. Heinrich, D.N. Christodoulidies, M. Khajavikan. Parity-time–symmetric microring lasers,
  \href{https://doi.org/10.1126/science.1258480}{Science \textbf{346}, 975 (2014).}

\bibitem{ref41}
  L. Feng, Z.J. Wong, R.-M. Ma, Y. Wang, X. Zhang. Single-mode laser by parity-time symmetry breaking,
  \href{https://doi.org/10.1126/science.1258479}{Science \textbf{346}, 972 (2014).}

\bibitem{ref42}
  S. De Liberato, D. Gerace, I. Carusotto, C. Ciuti. Extracavity quantum vacuum radiation from a single qubit,
  \href{https://doi.org/10.1103/PhysRevA.80.053810}{Phys. Rev. A \textbf{80}, 053810 (2009).}

\bibitem{ref43}
  F. Beaudoin, J.M. Gambetta, A. Blais. Dissipation and ultrastrong coupling in circuit QED,
  \href{https://doi.org/10.1103/PhysRevA.84.043832}{Phys. Rev. A
        \textbf{84}, 043832 (2011).}

\bibitem{ref44}
  A. Ridolfo, M. Leib, S. Savasta, M.J. Hartmann. Photon Blockade in the Ultrastrong Coupling Regime,
  \href{https://doi.org/10.1103/PhysRevLett.109.193602}{Phys. Rev. Lett. \textbf{109}, 193602 (2012).}

\bibitem{ref45}
  R. Stassi, S. Savasta, L. Garziano, B. Spagnolo, F. Nori. Output field-quadrature measurements and squeezing in ultrastrong cavity-QED,
  \href{https://doi.org/10.1088/1367-2630/18/12/123005}{New J. Phys. \textbf{18}, 123005 (2016).}

\bibitem{ref46}
  S. De Liberato. Virtual photons in the ground state of a dissipative system,
  \href{https://doi.org/10.1038/s41467-017-01504-5}{Nat. Commun. \textbf{8}, 1465 (2017).}

\bibitem{ref47}
  M. Bamba, T. Ogawa. Dissipation and detection of polaritons in the ultrastrong-coupling regime,
  \href{https://doi.org/10.1103/PhysRevA.86.063831}{Phys. Rev. A \textbf{86}, 063831 (2012).}

\bibitem{ref48}
  M. Bamba, T. Ogawa. System-environment coupling derived by Maxwell's boundary conditions from the weak to the ultrastrong light-matter-coupling regime,
  \href{https://doi.org/10.1103/PhysRevA.88.013814}{Phys. Rev. A \textbf{88}, 013814 (2013).}

\bibitem{ref49}
  M. Bamba, T. Ogawa. Recipe for the Hamiltonian of system-environment coupling applicable to the ultrastrong-light-matter-interaction regime,
  \href{https://doi.org/10.1103/PhysRevA.89.023817}{Phys. Rev. A \textbf{89}, 023817 (2014).}

\bibitem{ref50}
  S. De Liberato. Comment on “System-environment coupling derived by Maxwell's boundary conditions from the weak to the ultrastrong light-matter-coupling regime”,
  \href{https://doi.org/10.1103/PhysRevA.89.017801}{Phys. Rev. A \textbf{89}, 017801 (2014).}

\bibitem{ref51}
  M. Bamba, K. Inomata, Y. Nakamura. Superradiant Phase Transition in a Superconducting Circuit in Thermal Equilibrium,
  \href{https://doi.org/10.1103/PhysRevLett.117.173601}{Phys. Rev. Lett. \textbf{117}, 173601 (2016).}

\bibitem{ref52}
  V.Y. Shishkov, E.S. Andrianov, A.A. Pukhov, A.P. Vinogradov, A.A. Lisyansky. Relaxation of interacting open quantum systems,
  \href{https://doi.org/10.3367/UFNe.2018.06.038359}{Phys.-Uspekhi \textbf{62}, 510 (2019).}

\bibitem{ref53}
  V.Y. Shishkov, E.S. Andrianov, A.A. Pukhov, A.P. Vinogradov, A.A. Lisyansky. Perturbation theory for Lindblad superoperators for interacting open quantum systems,
  \href{https://doi.org/10.1103/PhysRevA.102.032207}{Phys. Rev. A \textbf{102}, 032207 (2020).}

\bibitem{ref54}
  R. Kosloff. Quantum Thermodynamics: A Dynamical Viewpoint,
  \href{https://doi.org/10.3390/e15062100}{Entropy \textbf{15}, 2100 (2013).}

\bibitem{ref55}
  D. Mozgunov, D. Lidar. Completely positive master equation for arbitrary driving and small level spacing,
  \href{https://doi.org/10.22331/q-2020-02-06-227}{Quantum \textbf{4}, 227 (2020).}

\bibitem{ref56}
  F. Nathan, M. S. Rudner. Universal Lindblad equation for open quantum systems,
  \href{https://doi.org/10.1103/PhysRevB.102.115109}{Phys. Rev. B \textbf{102}, 115109 (2020).}

\bibitem{ref57}
  S. Scali, J. Anders, L. A. Correa. Local master equations bypass the secular approximation, 
  \href{https://doi.org/10.22331/q-2021-05-01-451}{Quantum \textbf{5}, 451 (2021).}

\bibitem{ref77}
G. Hackenbroich, C. Viviescas, F. Haake. Quantum statistics of overlapping modes in open resonators, \href{https://doi.org/10.1103/PhysRevA.68.063805}{Phys. Rev. A \textbf{68}, 063805 (2003).}

\bibitem{ref58}
  I.V. Vovchenko, V.Y. Shishkov, A.A. Zyablovsky, E.S. Andrianov. Model for the description of the relaxation of quantum-mechanical systems with closely spaced energy levels,
  \href{https://doi.org/10.1134/S0021364021130099}{JETP Letters \textbf{114}, 51-57 (2021).}
  
\bibitem{ref71}
  H. Carmichael.
  \href{https://link.springer.com/book/10.1007/978-3-540-47620-7}{An open systems approach to quantum optics, Springer-Verlag, Berlin (1991).}
 
\bibitem{ref72}
  C.W. Gardiner, P. Zoller.
  \href{https://www.springer.com/gp/book/9783540223016?token=gbgen&wt_mc=GoogleBooks.GoogleBooks.3.EN}{Quantum noise: A handbook of Markovian and non-Markovian quantum stochastic methods with applications to quantum optics, Springer-Verlag, Berlin (2004).}

\bibitem{ref59}
  J.O. Gonzalez, L.A. Correa, G. Nocerino, J.P. Palao, D. Alonso, G. Adesso. Testing the validity of the 'local' and 'global' GKLS master equations on an exactly solvable model,
  \href{https://doi.org/10.1142/S1230161217400108}{Open Syst. Inf. Dyn. \textbf{24}, 1740010 (2017).}

\bibitem{ref60}
  A. Rivas, A.D.K. Plato, S.F. Huelga, M.B. Plenio. Markovian master equations: a critical study,
  \href{https://doi.org/10.1088/1367-2630/12/11/113032}{New J. Phys. \textbf{12}, 113032 (2010).}

\bibitem{ref78}
I. I. Arkhipov, A. Miranowicz, F. Minganti, F. Nori. Liouvillian exceptional points of any order in dissipative linear bosonic systems: Coherence functions and switching between PT and anti-PT symmetries, \href{https://doi.org/10.1103/PhysRevA.102.033715}{Phys. Rev. A \textbf{102}, 033715 (2020).}

\bibitem{ref79}
F. Minganti, A. Miranowicz, R. W. Chhajlany, F. Nori. Quantum exceptional points of non-Hermitian Hamiltonians and Liouvillians: The effects of quantum jumps, \href{https://doi.org/10.1103/PhysRevA.100.062131}{Phys. Rev. A \textbf{100}, 062131 (2019).}

\bibitem{ref80}
F. Minganti, A. Miranowicz, R. W. Chhajlany, I. I. Arkhipov, F. Nori. Hybrid-Liouvillian formalism connecting exceptional points of non-Hermitian Hamiltonians and Liouvillians via postselection of quantum trajectories, \href{https://doi.org/10.1103/PhysRevA.101.062112}{Phys. Rev. A \textbf{101}, 062112 (2020).}

\bibitem{ref81}
I. I. Arkhipov, A. Miranowicz, F. Minganti, F. Nori. Quantum and semiclassical exceptional points of a linear system of coupled cavities with losses and gain within the Scully-Lamb laser theory, \href{https://doi.org/10.1103/PhysRevA.101.013812}{Phys. Rev. A \textbf{101}, 013812 (2020).}

\bibitem{ref84}
S. Franke, S. Hughes, M. K. Dezfouli, P. T. Kristensen, K.
Busch, A. Knorr, M. Richter. Quantization of quasinormal modes for open cavities and plasmonic cavity quantum electrodynamics, \href{https://doi.org/10.1103/PhysRevLett.122.213901}{Phys. Rev. Lett. \textbf{122}, 213901 (2019).}

\bibitem{ref76}
 T. Pickering, J. M. Hamm, A. F. Page, S. Wuestner, O. Hess. (2014). Cavity-free plasmonic nanolasing enabled by dispersionless stopped light, \href{https://doi.org/10.1038/ncomms5972}{Nat. Commun. \textbf{5}, 4972 (2014).}
  
 \bibitem{ref61}
  A.F. Kockum, A. Miranowicz, S. De Liberato, S. Savasta, F. Nori. Ultrastrong coupling between light and matter,
  \href{https://doi.org/10.1038/s42254-018-0006-2}{Nature Reviews Physics \textbf{1}, 19-40 (2019).}

\bibitem{ref62}
  P. Forn-Díaz, L. Lamata, E. Rico, J. Kono, E. Solano. Ultrastrong coupling regimes of light-matter interaction,
  \href{https://doi.org/10.1103/RevModPhys.91.025005}{Rev. Mod. Phys. \textbf{91}, 025005 (2019).}

\bibitem{ref63}
  T. Niemczyk, F. Deppe, H. Huebl, et al. Circuit quantum electrodynamics in the ultrastrong-coupling regime,
  \href{https://doi.org/10.1038/nphys1730}{Nature Phys. \textbf{6}(10), 772-776 (2010).}
 
\bibitem{ref64}
  I.I. Rabi. Space quantization in a gyrating magnetic field,
  \href{https://doi.org/10.1103/PhysRev.51.652}{Phys. Rev.        \textbf{51}, 652 (1937).}

\bibitem{ref65}
  R.H. Dicke. Coherence in spontaneous radiation processes,
  \href{https://doi.org/10.1103/PhysRev.93.99}{Phys. Rev. A \textbf{93}, 99 (1954).}

\bibitem{ref66}
  A.A. Anappara, S. De Liberato, A. Tredicucci, C. Ciuti, G. Biasiol, L. Sorba, F. Beltram. Signatures of the ultrastrong light-matter coupling regime,
  \href{https://doi.org/10.1103/PhysRevB.79.201303}{Phys. Rev. B \textbf{79}, 201303 (2009).}

\bibitem{ref67}
  P. Forn-Díaz, J. Lisenfeld, D. Marcos, J.J. Garcia-Ripoll, E. Solano, C.J.P.M. Harmans, J.E. Mooij. Observation of the Bloch-Siegert shift in a qubit-oscillator system in the ultrastrong coupling regime,
  \href{https://doi.org/10.1103/PhysRevLett.105.237001}{Phys. Rev. Lett. \textbf{105}, 237001 (2010).}

\bibitem{ref68}
  C. Ciuti, G. Bastard, I. Carusotto. Quantum vacuum properties of the intersubband cavity polariton field,
  \href{https://doi.org/10.1103/PhysRevB.72.115303}{Phys. Rev. B \textbf{72}, 115303 (2005).}

\bibitem{ref69}
  X. Gu, A. F. Kockum, A. Miranowicz, Y.-X. Liu, F. Nori. Microwave photonics with superconducting quantum circuits,
  \href{https://doi.org/10.1016/j.physrep.2017.10.002}{Phys. Rep. \textbf{718-719}, 1-102 (2017).}

\bibitem{ref70}
  D. Z. Rossatto, C. J. Villas-Bôas, M. Sanz, E. Solano. Spectral classification of coupling regimes in the quantum Rabi model,
  \href{https://doi.org/10.1103/PhysRevA.96.013849}{Phys. Rev. A \textbf{96}, 013849 (2017).}

\bibitem{ref73}
  A. Balanov, N. Janson, D. Postnov, O. Sosnovtseva.
  \href{https://link.springer.com/book/10.1007/978-3-540-72128-4}{Synchronization: From simple to complex, Springer (2009).}

 \bibitem{ref74}
  J. Plemelj.
  \href{https://doi.org/10.1002/zamm.19650450117}{Problems in the sense of Riemann and Klein, Interscience Publishers, New York (1964).}

\bibitem{ref75}
  M. Cattaneo, G.L. Giorgi, S. Maniscalco, R. Zambrini. Local versus global master equation with common and separate baths: superiority of the global approach in partial secular approximation, \href{https://doi.org/10.1088/1367-2630/ab54ac}{New J. Phys. \textbf{21}, 113045 (2019).}

\end{thebibliography}

\end{document}